# Discovering Kinetically Significant Reaction Mechanisms Beyond Chemical Intuition in Condensed-Phase Radiolysis

*Nitesh Kumar,[a,b] Jacob R. Milton,[a,b] Eric Sivonxay,[d] Brett A. Helms,[a,e] Frances A. Houle*[a,c,f] and Samuel M. Blau*[a,d]*

a. Center for High Precision Patterning Science, Lawrence Berkeley National Laboratory, Berkeley, California 94720, United States

b. Materials Sciences Division, Lawrence Berkeley National Laboratory, Berkeley, California 94720, United States

c. Chemical Sciences Division, Lawrence Berkeley National Laboratory, Berkeley, California 94720, United States

d. Energy Technologies Area, Lawrence Berkeley National Laboratory, Berkeley, California 94720, United States

e. The Molecular Foundry, Lawrence Berkeley National Laboratory, Berkeley, California 94720, United States

f. Molecular Biophysics and Integrated Bioimaging Division, Lawrence Berkeley National Laboratory, Berkeley, California 94720, United States

Email: smblau@lbl.gov, fahoule@lbl.gov

## Abstract

Many important chemical systems — from radiation-driven processes to condensed-phase photochemistry — involve reaction mechanisms that are so complex it is a challenge to characterize them experimentally or predict them from chemical intuition. Existing computational approaches to mechanism discovery typically assess pathway importance through thermodynamic favorability alone, which does not provide time-dependent kinetics or inform how to model spatial inhomogeneities. Here we describe an integrated workflow that discovers complex reaction mechanisms without prescribing them and connects molecular-scale reactivity to spatiotemporal observables. The workflow combines high-throughput DFT, automated reaction network construction with chemical plausibility filtering, stochastic pathway sampling to identify reactions which are likely to occur, and spatially resolved reaction–diffusion kinetics simulations with explicit tracking of species in space and time. To demonstrate the workflow on a system of high complexity, we apply it to radiolytic chemistry in an extreme ultraviolet (EUV) organic polymer thin film photoresist, where a single 92 eV photon initiates cascades of radical ions, fragments, and low-energy electrons across a nanoscale radiolytic spur. Starting from over 3,300 species and millions of candidate reactions, the workflow identifies the most likely reaction pathways and produces spatiotemporal maps that resolve product formation on femtosecond-to-nanosecond timescales across a 15.5-nm domain. The simulations predict products detected experimentally and reveal that the identity of the initially photoionized species profoundly shapes the downstream product distribution through multi-step pathways governing the balance between deprotection and crosslinking reactions. The methodology is broadly applicable to complex condensed-phase reactive systems.

# Introduction

Many important chemical systems such as condensed-phase radiolysis, aerosol and thin-film photochemistry, and electrochemical interphase formation involve reaction mechanisms that are too complex and too incompletely characterized to be elucidated by experiment or affordable rigorous theory. Reaction mechanism discovery via high-throughput calculations has emerged as a powerful strategy for such systems, enabling systematic enumeration and analysis of large sets of candidate pathways with fewer assumptions than intuition-driven approaches or methods which depend on pre-set reaction class templates.[1-10] However, this methodology often produces reaction networks comprising tens of thousands to millions of steps, most of which are likely to be kinetically unimportant — i.e., not active in the overall product formation mechanism. Identifying the relevant pathways requires robust selection strategies that go beyond thermodynamic favorability, which on its own, provides neither time-dependent kinetics information nor a basis for modeling spatial inhomogeneities that are ubiquitous in real systems.

To address these limitations, we developed a workflow that connects molecular-scale reactivity to spatiotemporal observables without prescribing specific mechanisms. The computational workflow combines density functional theory (DFT) optimization of molecular species, automated construction of balanced chemical reactions filtered by physics-based plausibility criteria, stochastic sampling of reaction sequences to identify kinetically relevant pathways, and spatially resolved stochastic kinetic simulation that tracks product formation in both space and time.[11-13] The stages of this pipeline are interdependent. Enumeration alone produces an intractable reaction space, filtering yields a tractable network but does not establish which reactions are most likely to occur, and pathway sampling identifies the kinetically relevant reactions but not where and when products form. Integrating these stages, each informed by first-principles energetics and physics- and chemistry-based kinetics, enables mechanism discovery in systems where chemical intuition alone would miss essential pathways, and provides a means of direct comparison to spatiotemporal experimental observations.

To evaluate this workflow on a system of genuine complexity, we apply it to a specific case of condensed-phase chemistry: light-driven reactions initiated within polymer films by extreme ultraviolet (EUV) photons during photolithographic patterning for semiconductor manufacturing.[14, 15] EUV lithography uses 13.5 nm photons (~92 eV) to drive radiolytic processes that are fundamentally different from thermal or direct photochemical reactions, producing radical ions, energetic electrons, and electronically excited fragments whose subsequent chemistry is governed by energy deposition rather than barrier heights — a picture consistent with the negligible temperature dependence long established for primary radiolytic processes.[16-18] This system represents an ideal test case: the mechanisms are incompletely characterized, the chemistry is too complex for intuitive prediction, spatial and temporal inhomogeneity at the molecular scale strongly shapes the resulting chemistry, and several experimentally detected products are available for validation.

Understanding the specific challenges of EUV-driven chemistry requires some background on photolithographic imaging. In photolithography, a photoresist — a thin photosensitive polymer film — is exposed to patterned radiation to form a latent chemical image, which is then converted to a developable image by a post-exposure bake (~100°C for 1 minute) and developed in aqueous base to produce topographic patterns suitable for semiconductor fabrication.[15] Modern chemically amplified resists contain a polymer, a photoacid generator, and a base quencher. Under deep ultraviolet (DUV) illumination at 248 nm or 193 nm, the photoacid generator selectively undergoes well-localized singlet or triplet excitations that produce a strong

Brønsted acid,[19-21] which then catalyzes numerous deprotection reactions during the bake step. Under EUV, the physics changes qualitatively. Each 13.5 nm photon deposits ~92 eV into the film — enough to photoionize any molecular species present, precluding selective activation of a single film component. All components of the resist can absorb and react, direct acid generation is not the dominant process,[22, 23] and the resulting chemistry is driven by radical cations, radical anions, free radicals, and electronically excited fragments that each open reaction channels inaccessible at longer wavelengths. Each photoabsorption event generates on average three secondary electrons in addition to the initial photoelectron.[24] Species that do not directly absorb an EUV photon can nonetheless dominate downstream chemistry through their affinity for secondary electrons.

These features make EUV resist patterning chemistry a compelling but challenging target for computational modeling. Prior approaches to simulate this process include electron scattering models,[22, 25] Monte Carlo simulations,[26, 27] continuum reaction–diffusion frameworks,[28] stochastic chemical kinetics,[29] and blended multiscale methods.[30] These studies have generated important mechanistic insights; however, a detailed connection between radiation physics and the chemistry of fully formulated resist systems remains incomplete, largely because of the lack of information on probable reaction pathways, their timing, and the spatial localization of individual events — despite notable recent advances.[31] Spatial resolution is critical because the spatial characteristics of the initial latent image directly determine both the achievable resolution of the patterned film and the outcomes of the subsequent bake process, which has been characterized for DUV resists[32] but remains poorly understood for EUV.

The study described here demonstrates that it is possible to computationally access detailed mechanistic information about how complex condensed-phase reactions evolve in space and time at the molecular level, with the specific example of EUV-initiated radiolytic chemistry in a chemically amplified photoresist. By resolving both kinetics and spatial–temporal evolution across the nanoscale radiolytic spur, our methodology provides a mechanistically grounded view of how reactive intermediates form, migrate, and produce the chemical changes central to photolithographic latent image formation — insights that are inaccessible to any single computational approach in isolation and that connect directly to experimental detection of reaction products. More broadly, the workflow established here is applicable to any complex condensed-phase system in which reaction mechanisms are incompletely characterized and spatiotemporal resolution is needed to connect sequences of molecular-scale events to averaged macroscopic observables.

## Methodology

Connecting molecular-scale reactivity to the spatiotemporal evolution of a complex condensed-phase system requires bridging a vast knowledge gap between what is chemically possible and what actually occurs. Our approach (**Figure 1**) addresses this by systematically constructing the space of plausible molecular species and reactions without prescribing specific mechanisms, then progressively filtering that space through automated chemical logic and stochastic pathway sampling, and finally embedding the surviving reactions in a spatially resolved kinetic simulation that tracks product formation in both space and time. The result is a workflow that can discover reaction mechanisms that would not be identified by intuition-driven or template-based approaches, while remaining grounded in first-principles energetics and physics and chemistry-based kinetics. We describe each stage below, following the workflow shown in **Figure 1**.

The model system for this study is a part of the Environmentally Stable Chemically Amplified Photoresist (ESCAP) formulation class.[32] The photoresist comprises five components: a 60:40 random copolymer of poly(4-hydroxystyrene) (PHS) and poly(*tert*-butylmethacrylate) (PtBMA), with triphenylsulfonium perfluorobutane sulfonate, also called nonaflate, ($TPS^+$ $Nf^-$) and 4-cyanobenzoate ($TPS^+$ $CNBZ^-$) as additives to control image formation (**Figure 1a**). The latter species are respectively the photoacid generator and base quencher when DUV light is used; as noted they do not have this same function in EUV, so we will only refer to them through their chemical identities. Pendant functional groups along polymer chains are the primary chemically active sites during exposure. The polymer backbone is represented using saturated hydrocarbon segments that retain enough of the chain structure to preserve the electronic environment of the reactive functional groups, while avoiding artifacts from chain truncation. Although we demonstrate the workflow on this specific material, it is general and applicable to other complex reactive systems driven by photoionization.

A reaction network for this photoresist composition (**Figure 1b**) is constructed in three stages. The first stage of the workflow establishes the full set of molecular species that could plausibly participate in the chemistry (**Figure 1c**). Species fall into three classes: the five starting components, molecular fragments derived from bond fragmentation of those components, and recombinant species formed by combining fragments or starting species into new molecular structures. A fragment in this set makes no claim that it is physically formed as a discrete intermediate during EUV exposure, since many recombinants arise through concerted processes in which the fragment never exists independently. Fragment generation proceeds by iteratively cleaving each symmetry-unique single bond in every starting species, subject to two constraints. Backbone bond cleavage within the PHS and PtBMA monomers was excluded because recombinants built from backbone fragments lack a plausible formation pathway in this system: strong cage effects would prevent the spatial separation needed for any such recombinant to form.[33, 34] Fragmentation of the nonaflate anion was restricted to cleavage of the carbon–sulfur bond. Permitting carbon–fluorine cleavage — where bond dissociation energies are substantially higher — generated a large set of recombinants with no chemically reasonable route to formation. Carbon–carbon bond cleavage was allowed but the resulting fragments were not included as recombinants. These constraints yielded 108 unique fragment species. Recombinant species were then generated by combinatorially merging fragments and starting species using open-source cheminformatics tools.[35-37] To focus the recombinant set on chemically realistic combinations and keep the subsequent DFT calculations tractable, we introduced bonding rules derived from local electronic structure: a new bond was permitted only between an undercoordinated atom on one species and either an undercoordinated or radical-bearing site on the other, as determined by standard valence rules and spin density analysis. This produced 3,253 recombinant species. Together with the 108 fragments and 5 starting species, the complete species set defines the chemical space from which all candidate reactions are drawn.

Free energies for every species were computed via geometry optimization and vibrational frequency calculations using high-throughput DFT workflows[38] that automate error handling and ensure consistent treatment across thousands of structurally diverse molecules. Computational details including the density functional, basis set, and implicit solvation model are provided in **Supporting Information (SI) section S1a.**

From these species and their computed energetics, we next generated reactions connecting them via combinatorial enumeration followed by filtering using chemical logic (**Figure 1d**). The number of stoichiometrically balanced reactions that can be written among

~3,400 species is on the order of hundreds of millions. HiPRGen[39] reduces this space by applying a series of rule-based filters that enforce chemical plausibility — e.g., charge and spin conservation, a realistic number of bonds breaking and forming simultaneously, steric viability — producing approximately 2.9 million candidate reactions that survive these checks. The filters for each stage are designed to be few in number, general, and physically justifiable — they exclude reactions that cannot reasonably occur rather than prescribing those that should, preserving a far broader reaction space than reaction-template-based enumeration would access. The detailed filter criteria for each stage are described in **SI Section S1b.**

The filtering proceeded in two stages that reflect the distinct physical regimes of EUV-initiated chemistry. The first stage captured reactions associated with radiolytic activation: the processes driven by the ~92 eV deposited by an EUV photon. These include photoionization, fragmentations of radical cations and electronically excited species, and endergonic transformations accessible via the deposited energy. Electron attachment was also captured in this stage; although exergonic, these processes are effectively barrierless and occur as immediate consequences of the ionization events that generate free electrons, making them inseparable from the activation chemistry. The second stage captured the subsequent exergonic chemistry — small fragment transfer, radical recombination, and molecular rearrangement — through which the reactive intermediates produced during activation evolve toward stable products.

While the 2.9 million reactions that survive HiPRGen's chemical plausibility filters are all individually reasonable, the vast majority will not participate meaningfully in the chemistry of any given system if the relevant reactants are never formed. To identify the most relevant pathways, we coupled HiPRGen with Reaction Network Monte Carlo (RNMC),[39] which stochastically samples the filtered network to determine which reactions occur and how frequently (**Figure 1e**). RNMC uses a Monte Carlo framework to propagate the reaction network from a defined starting state, tallying reaction events across many independent trajectories. For each stage, 100,000 trajectories were run; the terminal states of the activation-stage trajectories served as the starting conditions for the exergonic-stage trajectories. Reactions that occurred fewer than 500 times across all trajectories were discarded. It is important to note that this pathway sampling step operates under simplified conditions — approximate rate coefficients, a well-mixed assumption, and no explicit spatial resolution — and is not intended to produce physically accurate kinetics. Its purpose is solely to identify the most-trafficked pathways from among the millions of candidates, so that the subsequent kinetic simulation becomes computationally tractable. System-specific corrections applied during this step are described in **SI Section S1b.** The combined HiPRGen filtering and RNMC pathway sampling reduced the reaction space to approximately 8,300 reactions.

These reactions, together with 47 EUV radiolysis steps established in our previously published model (**Figure 1f**),[40] were then incorporated into the stochastic kinetics platform Kinetiscope[11] (**SI Section S1c**) for physically grounded reaction–diffusion simulation (**SI section S2**) of the radiolytic spur created by absorption of a single EUV photon (**Figure 1g**). In contrast to the pathway sampling step, Kinetiscope propagated the full reaction–diffusion kinetics by tracking species concentrations with spatial resolution across the simulation domain and explicit electron energy tracking across discrete energy classes, all using physically grounded rate coefficients. Rate constants for the radiolysis steps were taken from our prior work.[40] For all other filtered chemical reactions, we assigned a uniform first order rate coefficient of $k = k_B T/h$, where $k_B$ is the Boltzmann constant, T is the absolute temperature, and h is the Planck constant. The species that drive spur formation chemistry — radical cations, electronically excited fragments, energetic electrons — carry energy that vastly exceeds typical reaction barriers, so barrier heights do not meaningfully limit their reaction rates. This is consistent with the negligible temperature dependence long established for primary radiolytic processes in condensed phases, where apparent activation energies are well below typical DFT accuracy thresholds.[16-18]

The kinetics simulation framework is a one-dimensional array of 31 compartments, each 0.5 nm deep, spanning a total spur length of 15.5 nm oriented in some direction within the uniform photoresist film (**Figure 1g**). Initial compositions in each compartment are determined by the resist film formulation, with species populations represented as discrete particles proportional to the component concentrations, 1 particle per molecular moiety, or 24 particles per compartment. A single 92 eV EUV photon is deposited into the first compartment, triggering photoabsorption and the subsequent radiolysis cascade. Electron transport is treated explicitly: high-energy electrons (~80, 55, and 30 eV) diffuse away from the absorption site, undergoing successive ionization and excitation events, progressively generating low-energy electrons (LEE, ~5 eV) and thermalized electrons (TE, ~0.1 eV) that diffuse bidirectionally between compartments. Molecular moieties do not diffuse, as expected for an ambient temperature solid-state resist environment over the ultrafast radiolysis timescale, so spatial evolution arises solely from electron migration and the location-specific chemistry that electrons induce. Each simulation proceeds until all reaction and diffusion probabilities decay to zero, ensuring that the product distributions capture the complete chemical evolution from radiolytic activation through final product formation. The exact sequence of reactions that occur is stochastic, with reaction event selection being driven by a random number generator initiated using a random number seed, and internal generation of the current time step associated with that event. 25 separate simulations of five cases were performed and their results averaged to discern trends among the spatiotemporal fluctuations. The five cases were photoexcitation of each of the five resist components shown in **Figure 1**, enabling an assessment of how the exposure chemistry depends on the specific photoabsorber.

The output of these simulations is a set of spatiotemporal maps that resolve product formation across the spur — what species form, where they form relative to the photoabsorption site (**SI Section S3**), and when they form on the femtosecond-to-nanosecond timescale of latent image formation (**Figure 1h**). This spatiotemporal resolution is what connects the molecular-scale reaction network to image formation, the first step toward the pattern-scale that ultimately determines photoresist performance, and it is not accessible through any of the individual workflow components in isolation. More generally, only by coupling comprehensive reaction enumeration with filtering by pathway sampling and spatially resolved stochastic kinetics can we propose and evaluate mechanisms for reactions of this complexity.

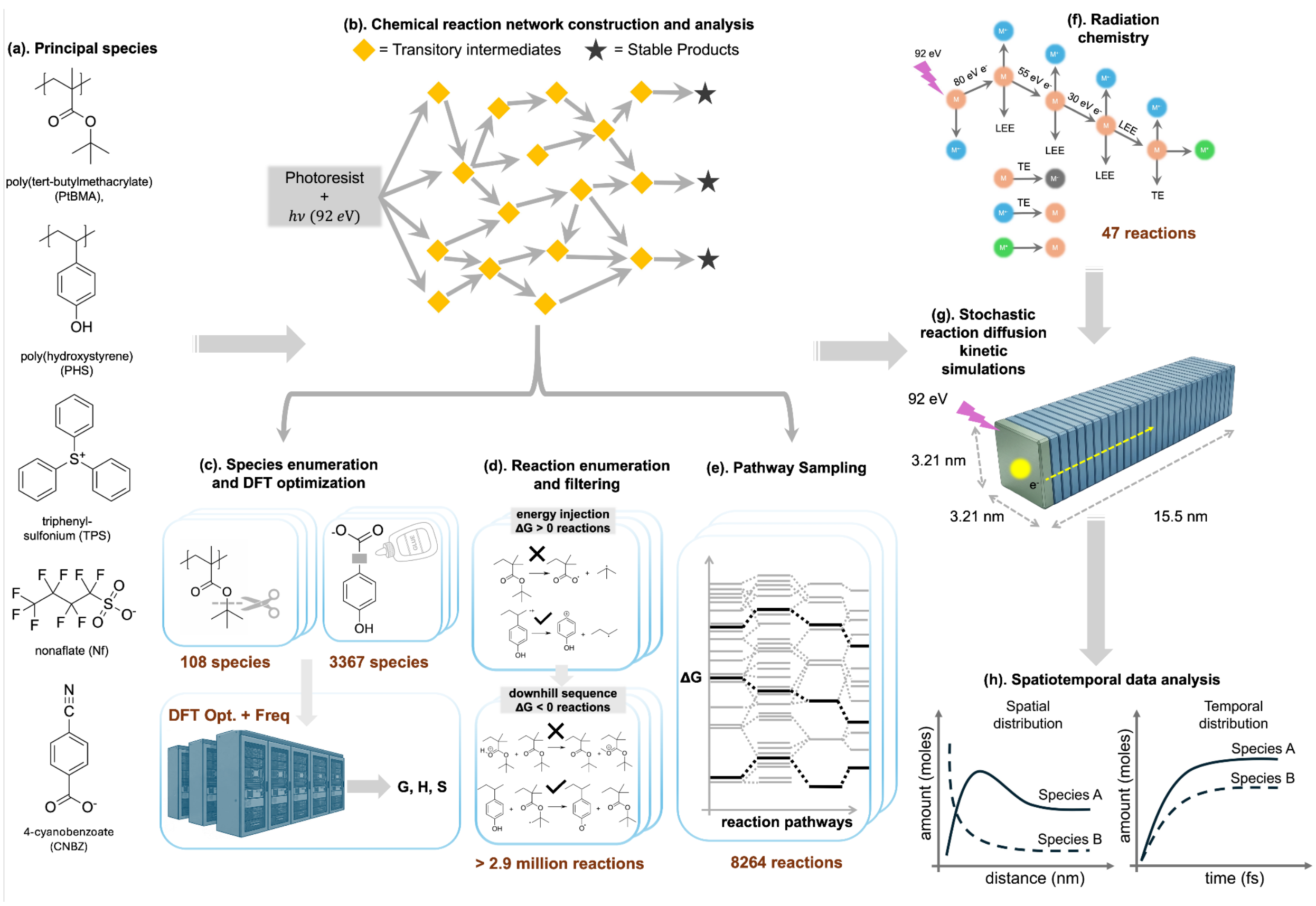

**Figure 1.** Schematic of the data-driven workflow that integrates high-throughput quantum chemistry, reaction network generation, filtering, and stochastic reaction-diffusion simulations for an EUV-exposed ESCAP-class photoresist. **(a)** Principal molecular components: poly(*tert*-butylmethacrylate) (PtBMA) and poly(4-hydroxystyrene) (PHS) monomers, triphenylsulfonium (TPS)-nonaflate (Nf), and TPS-4-cyanobenzoate (CNBZ). Their concentrations are 9.65 mol/L PHS, 7.29 mol/L PtBMA, 1.90 mol/L TPS-Nf, 2.39 mol/L TPS-CNBZ. The chemical reaction network **(b)** is then constructed through three steps: **(c)** The principal species and their fragments and recombinants yield 3367 unique species (including the electron); all species are subjected to DFT geometry optimization and frequency calculations to obtain thermodynamic quantities. **(d)** Reaction enumeration and filtering with HiPRGen yields over 2.9 million candidate reactions from the species pool. **(e)** Stochastic pathway sampling with Reaction Network Monte Carlo (RNMC) further down-selects to 8264 reactions of most significance. **(f)** The radiation chemistry initiated by a 92 eV EUV photon: successive inelastic scattering events produce secondary electrons at approximately 80, 55, and 30 eV, which further thermalize to low-energy electrons (LEE) and thermal electrons (TE), while generating electronically excited molecular states ($M^*$); this is encoded as 47 radiation-chemical initiation reactions. **(g)** The filtered chemical reaction network and the radiation-chemical initiation reactions are coupled into stochastic reaction-diffusion kinetic simulations performed over a discretized one-dimensional domain of 31 compartments with a total volume of 3.21 nm × 3.21 nm × 15.5 nm, where each compartment has a volume of 5.15 $nm^3$. **(h)** Data analysis of the simulation output yields spatial distributions of reaction products relative to the EUV interaction site and temporal profiles of species formation spanning femtosecond to nanosecond timescales.

## Results and Discussion

The chemistry driving pattern formation in EUV-exposed photoresists involves reactive processes too numerous and interconnected to be identified by chemical intuition or experiment alone, motivating this study's first-principles approach to reaction mechanism discovery. Herein, simulations of the physical radiolytic processes and the chemical processes defined by reaction network generation provide a detailed picture of events occurring in the film following EUV photoabsorption. The initial ionization, electron trapping and excitation events are coupled to formation of products by energetically downhill reactions, allowing the system to evolve beyond the initial radiolytic spur toward a fully reacted chemical state.

A single photoabsorption event generates about four ionization events contributing to subsequent chemistry leading to latent image formation.[24] The primary instigators for all chemical processes are the energetic electron (80 eV, 55 eV, 30 eV, all capable of ionization), low-energy electron (LEE, ~5 eV) and thermal electron (TE, ~0.1 eV) populations in the film. Spatiotemporal distributions of electrons combining results from all simulations are shown in **Figure 2**.

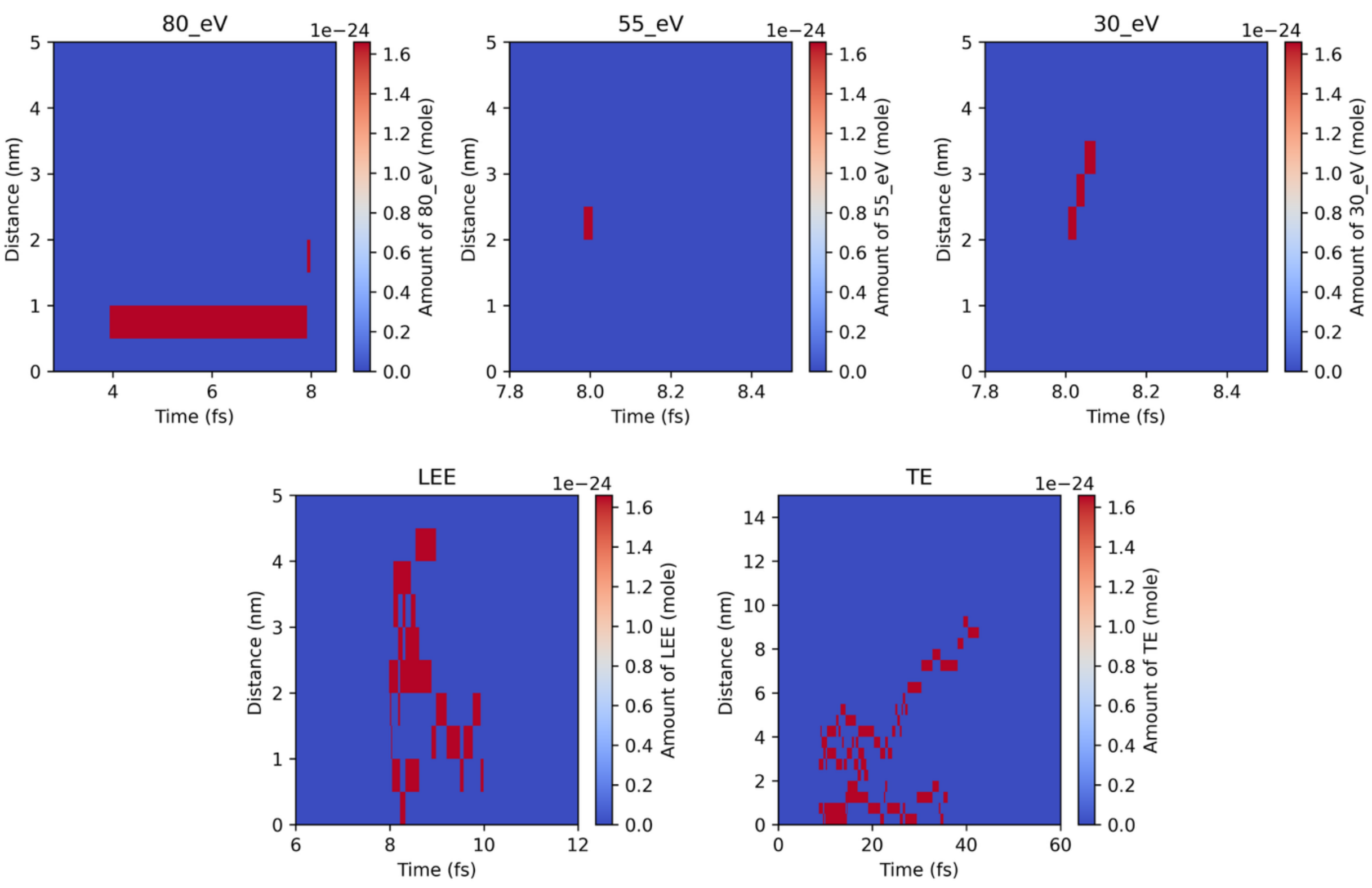

**Figure 2.** Spatiotemporal distributions of 80 eV, 55 eV, 30 eV, low-energy electrons (LEE) and thermal electrons (TE) — formed after initial event of 92 eV photon absorbance — plotted as amount versus time (fs) and distance (nm). The primary ionization event location is the compartment spanning 0–0.5 nm, the other compartments extend to the outermost location where EUV-absorption-generated electrons are found. Note that plot axes are at different scales.

In our simulations, highly energetic electrons (80, 55 and 30 eV) only forward scatter away from the absorption site, while all LEEs and TEs can diffuse both away from and toward the primary EUV absorption event location. As secondary electrons lose energy, their lifetimes increase, facilitating longer-range migration. This is because higher energy electrons drive ionization events which have high cross sections, while lower energy electrons can only be lost by electronic excitation and trapping events, which are relatively slow because of low LEE and TE concentrations (**Figure 2**). Electrons of all energies are concentrated within a nanoscale region around the absorption site, with a characteristic footprint of a few nanometers and a weak tail extending to 8–10 nm. This confinement is consistent with prior EUV spur modeling and Monte Carlo energy-loss simulations, which predict most-probable thermalized-electron travel distances of 3–5 nm with distributions extending to 8–10 nm, as well as extents of 6–7 nm for electron-driven product formation and a maximum reach of 10 nm.[22, 23, 40]

The spatially confined electron distribution initiates latent-image formation by generating reactive products near the photoabsorption site. These products provide a direct link between the initial electron formation and the chemical transformations that ultimately control pattern formation. Depending on the resist chemistry, subsequent reactions may involve deprotection, such as conversion of tert-butyl ester groups to carboxylic acid groups in PtBMA, crosslinking through covalent coupling between polymer fragments or chains, or other polarity-changing transformations that help determine resist tone (positive tone: exposed area dissolves during development, negative tone: unexposed area dissolves during development), with each pathway producing distinct molecular products and spatial distributions. Since highly energetic electrons are largely consumed at the site of photon absorption, the products formed in that compartment successfully capture the early steps of latent image formation, and are discussed below.

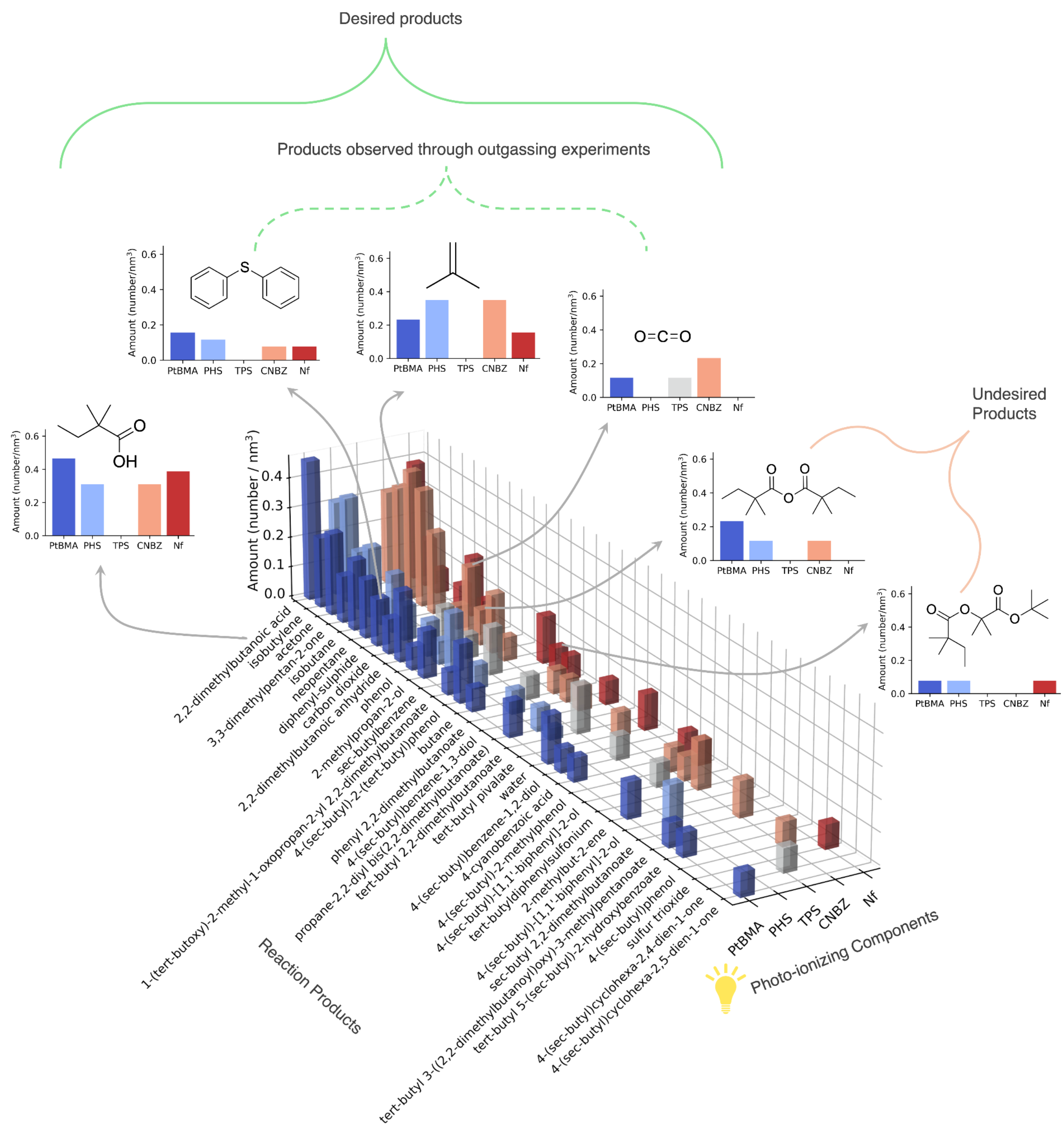

**Figure 3.** Concentrations of reaction products formed from a single photoabsorption event within a 5.15 nm$^3$ compartment volume, with the distributions of the most abundant species formed from each of the five principal species composing the photoresist material (defined in **Figure 1**). Illustrative structures for selected products are shown in the insets, which compare their abundances originating from photoexcitation of each resist component. Amounts are reported as averages over 25 independent simulations.

In our simulations, the photoabsorption event can be assigned to a specific resist component, namely, PHS and PtBMA, TPS cation, Nf anion, and CNBZ anion. The 30 most abundant reaction products formed in the photoexcited compartment initiated by each of the different photoabsorbers are presented in Figure 3. It is notable that direct photoabsorption by the TPS cation results in minimal reactivity within the 5.15 $nm^3$ resist photoabsorption region. The experimental information available on products formed are major signals observed by mass spectrometry due to outgassing from similar resist films during EUV exposure: isobutylene,[31, 41-44] diphenylsulfide,[45] acetone, carbon disulfide, toluene, *tert*-butyl benzene,[41] benzene,[43, 44] butane, $C_4H_6O$,[43] and $CO_2$.[31, 41-43] Many of these are included in Figure 3, but not carbon disulfide, benzene, *tert*-butyl benzene, $C_4H_6O$ and toluene. Examination of adjacent compartments reveals that, unlike aliphatic molecules like isobutylene, the aromatic molecules such as benzene or *tert*-butyl benzene are only formed outside of the photoexcitation volume (**see SI section S4**). This spatially resolved view of where individual products with different aromaticity originate is uniquely accessible through our simulation approach and, to our knowledge, has not been elucidated previously.

Although the resist material used in this study is meant to be positive tone, it is predicted to have some negative tone characteristics such as the formation of cross-linking reaction products, in agreement with a study of a similar resist system under EUV illumination which identified a conversion from positive to negative tone at high doses.[46] If the crosslinking persists during post-exposure bake, it would be expected to significantly influence local resist dissolution to form patterns[47] and therefore image characteristics.

The spatiotemporal distributions of deprotected PtBMA and crosslinked PtBMA provide examples of deprotection and crosslinking processes resulting in non-volatile photoresist film products, and are shown in **Figure 4** for all five photoabsorption cases (spatiotemporal distributions of a related volatile deprotection product, isobutylene, are shown in **Figure S1**). Deprotected PtBMA forms well beyond the initial absorption site, with appreciable amounts detected up to approximately ten nanometers from the initial light polymer interaction region. In contrast, the crosslinked product is more highly localized, largely confined to within 5 nm of the photon polymer interaction region. This spatial heterogeneity reflects how the differences in diffusion ranges of highly energetic electrons, LEEs and TEs, as shown in **Figure 2**, lead to differing compositions in spatially distinct regions of the film.

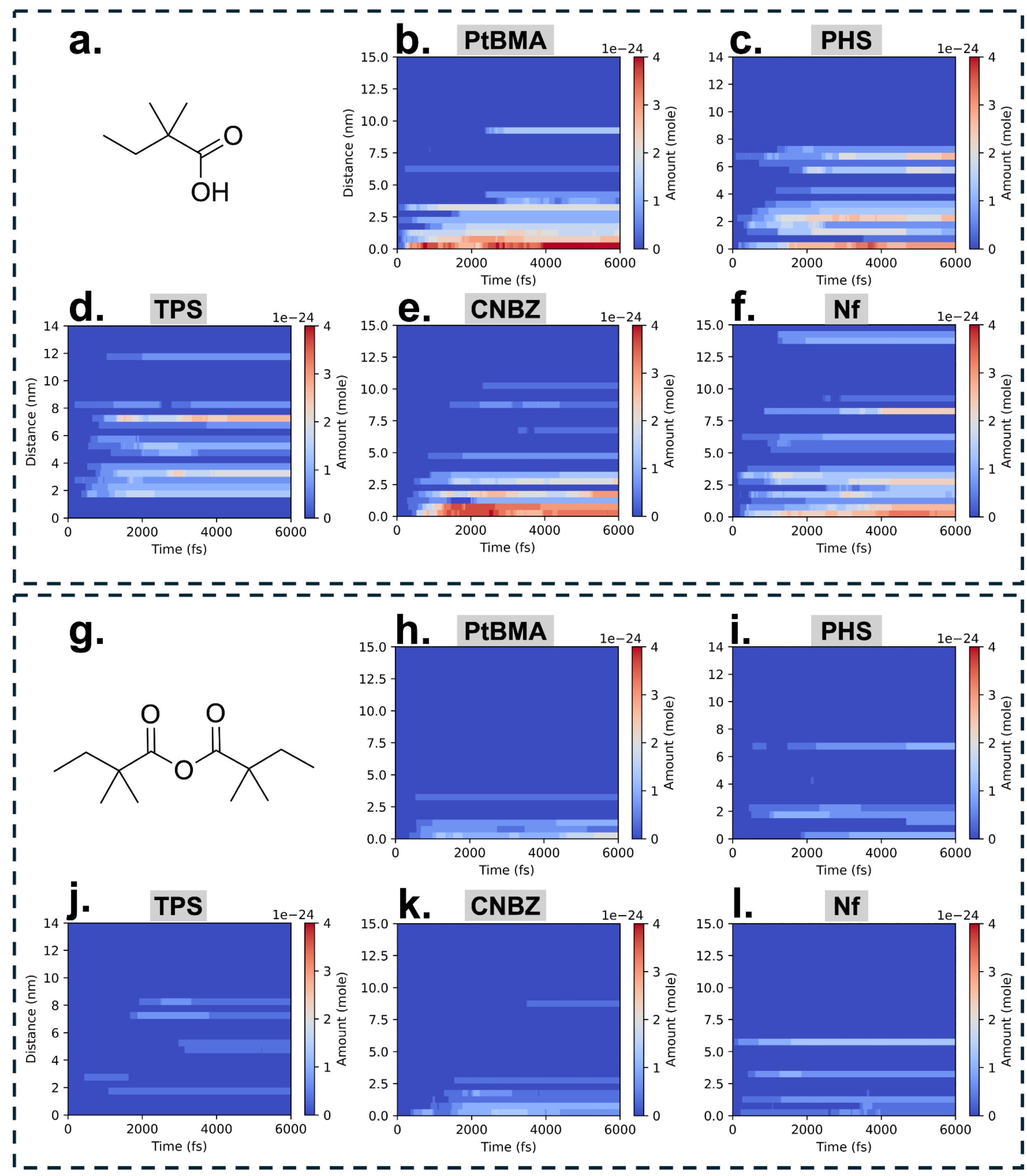

**Figure 4.** Spatiotemporal distributions of deprotection product 2,2-dimethylbutanoic acid and crosslinked product 2,2-dimethylbutanoic anhydride for each initially excited component (PtBMA, PHS, TPS, CNBZ, Nf respectively) of the photoresist, plotted as amount (moles, where 1.66 x $10^{-24}$ mole is equal to one molecule) versus time (up to 6000 fs) and distance (nm). Amounts are reported as averages over 25 independent simulations.

It is instructive to examine the spatiotemporal details of two example products — the deprotection product 2,2-dimethylbutanoic acid and the crosslinked product 2,2-dimethylbutanoic anhydride — as a function of the identity of the photoabsorber. When PtBMA absorbs, the primary deprotection product 2,2-dimethylbutanoic acid (**Figure 4a,b**) is found out to about 10 nm. The crosslinked product 2,2-dimethylbutanoic anhydride (**Figure 4g,h**) is confined to the region much closer to the absorption site (within 3 nm). When PHS is the primary absorber, 2,2-dimethylbutanoic acid (**Figure 4c**) is located within 8 nm of the absorption site, with the crosslinked product found out to 6 nm (**Figure 4i**). For TPS absorption-initiated events (**Figure 4 d,j**), product formation is displaced from the initial interaction site, reaching the maximum concentrations around 2 nm away from the absorption site. When CNBZ absorbs the EUV photon (**Figure 4e,k**), 2,2-dimethylbutanoic acid is found out to 10 nm, while the crosslinked product is more localized, similar to the trend found for PtBMA. Finally, when there is photoabsorption by the nonaflate anion Nf (**Figure 4f,l**), 2,2-dimethylbutanoic acid is found out to a much longer distance, 15 nm, with crosslinking occurring within 6 nm. Across all five photoabsorbers, deprotection reactions dominate over crosslinked network forming reactions. Additional analysis showing the spatiotemporal distribution of isobutylene, formed primarily via PtBMA deprotection reaction, is provided in **Figure S1** of the Supporting Information. While the patterns for isobutylene are fairly similar to the primary deprotection product, as one would expect, isobutylene can further undergo subsequent reactions in the film, so its amount at the end of the reaction-diffusion simulation is lower than that of the deprotected PtBMA product.

We further examined the time evolution of isobutylene, deprotected PtBMA and crosslinked PtBMA products, following photoabsorption by PtBMA in the 0–0.5 nm region of the resist (**Figure 5**), to elucidate how product formation progresses during latent image formation. The concentrations of isobutylene and 2,2-dimethylbutanoic acid increase rapidly for the first few 100 fs, then increase more slowly with competitive formation and consumption to an elapsed time of about 5000 fs. Crosslinked product formation rises more slowly until about 1000 fs, then continues to increase gradually to 5000 fs. The temporal evolution of the three PtBMA derived fragments within the first 0.5 nm compartment reveals the characteristic ordering of molecular events initiated immediately after EUV excitation. Isobutylene and 2,2 dimethylbutanoic acid are the first to form via bond cleavage, while formation of the crosslinked 2,2 dimethylbutanoic anhydride product proceeds more slowly and remains at lower abundance because it involves condensation of two fragments.

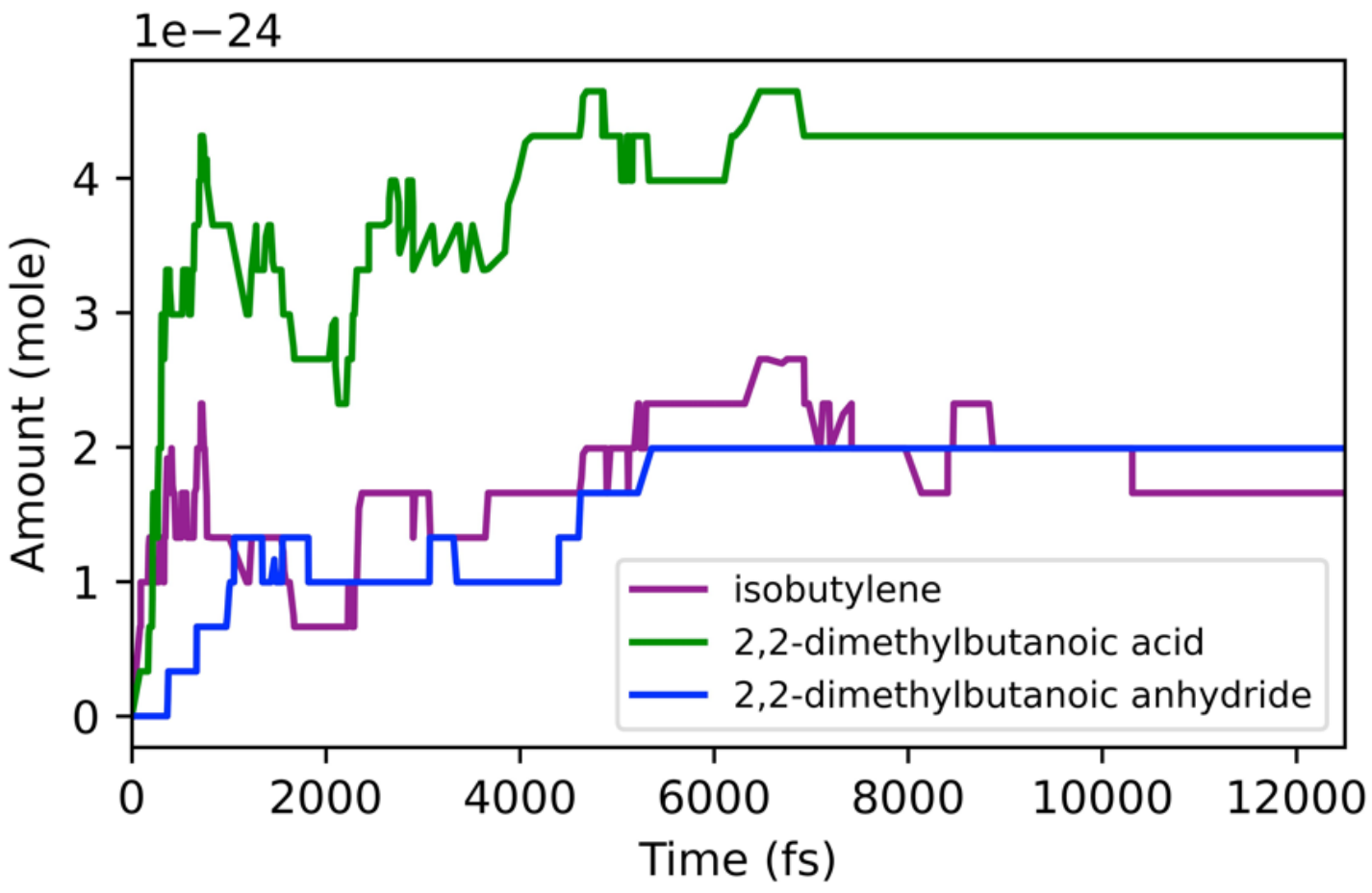

**Figure 5.** Temporal evolution of deprotection and crosslinked products following PtBMA exposure. Amounts are reported as averages over 25 independent simulations.

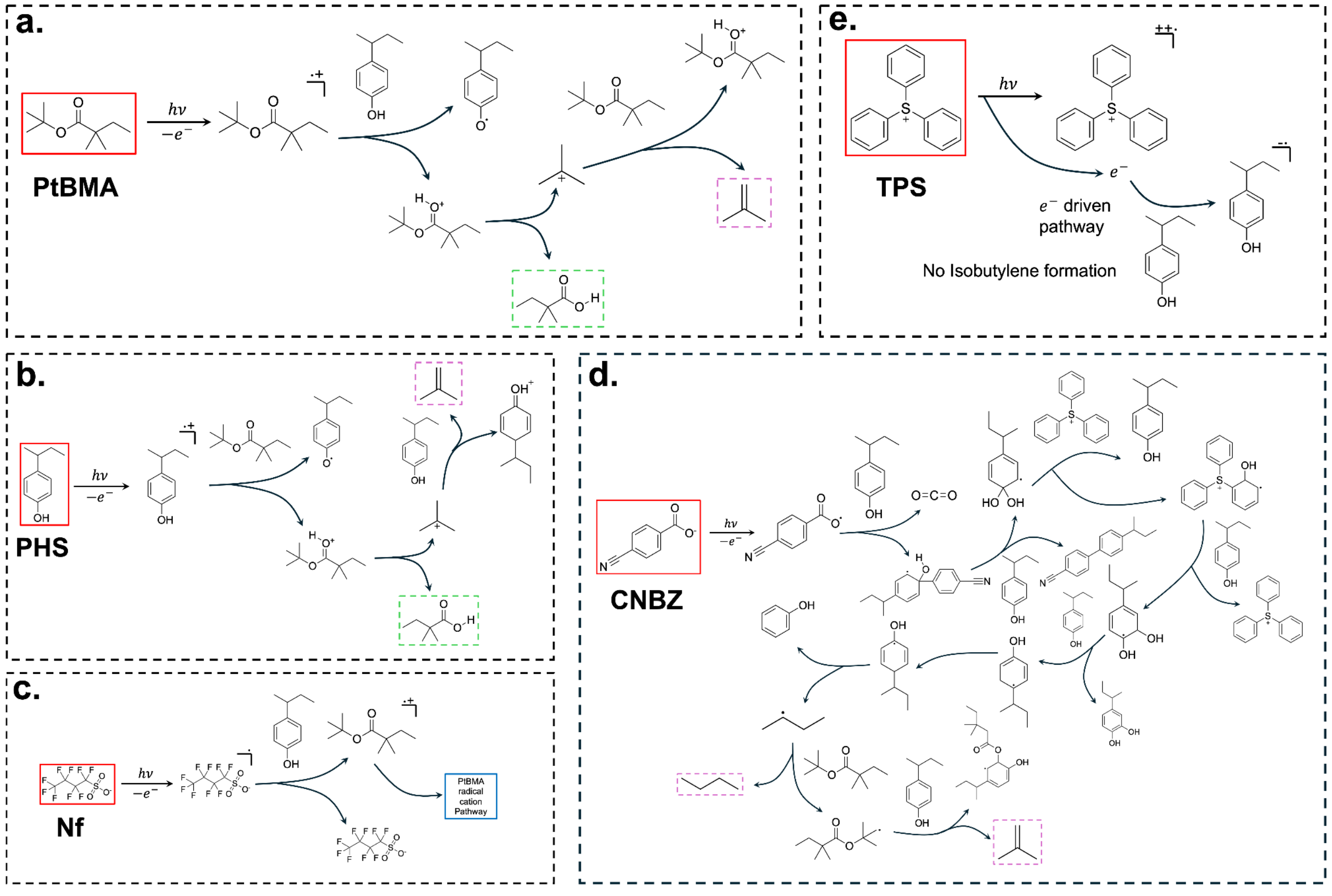

**Figure 6.** Mechanistic representation of the principal pathways that lead to isobutylene formation, grouped according to the identity of the molecule that first absorbs the EUV photon: (a) PtBMA, (b) PHS, (c) Nf, (d) CNBZ, and (e) TPS. Direct excitation of TPS (e) does not lead to local isobutylene formation.

Next, we tracked the time evolution of reaction sequences to highlight the coupling between the identity of the moiety that undergoes the initial photoabsorption event and the reaction pathways that originate thereafter. We specifically examine pathways responsible for the formation of isobutylene within the 5.15 $nm^3$ photoabsorption compartment (**Figure 6**) given its ubiquity in outgassing measurements[31, 41-44] for this and related photoresist systems, and because its formation is expected to result specifically from the deprotection event responsible for pattern formation**.** When PtBMA absorbs the incident photon, a radical cation is formed. This species can interact with an adjacent PHS moiety to abstract an H atom, forming a phenoxy radical and a protonated PtBMA that loses a *tert*-butyl cation to form 2,2-dimethylbutanoic acid deprotected product. The *tert*-butyl cation can transfer its proton to form isobutylene. When PHS is the initial absorber, a radical cation is formed that transfers a proton to an adjacent PtBMA, forming products by a pathway very similar to that for direct PtBMA excitation. The excitation of Nf generates a radical that rapidly abstracts an electron from the polymeric surroundings, promoting formation of the *tert*-butyl cation and its conjugate acid similar to direct polymer excitation. In contrast, direct excitation of CNBZ follows a largely radical driven route initiated by electron loss. Fragmentation coupled with decarboxylation and C-C bond scission directly generates isobutylene together with $CO_2$ without passing through a strongly acid mediated *tert*-butyl cation pathway. The free radical mechanism that yields isobutylene also results in crosslinking. TPS behaves differently from all of these cases. Direct excitation of TPS generates a di-cation that eventually serves as an electron trap but does not lead to local chemistry. The highly energetic electrons that are generated migrate into the surrounding polymer matrix, where they subsequently activate reactions involving the other resist components that follow their respective ionic or radical sequences to isobutylene formation.

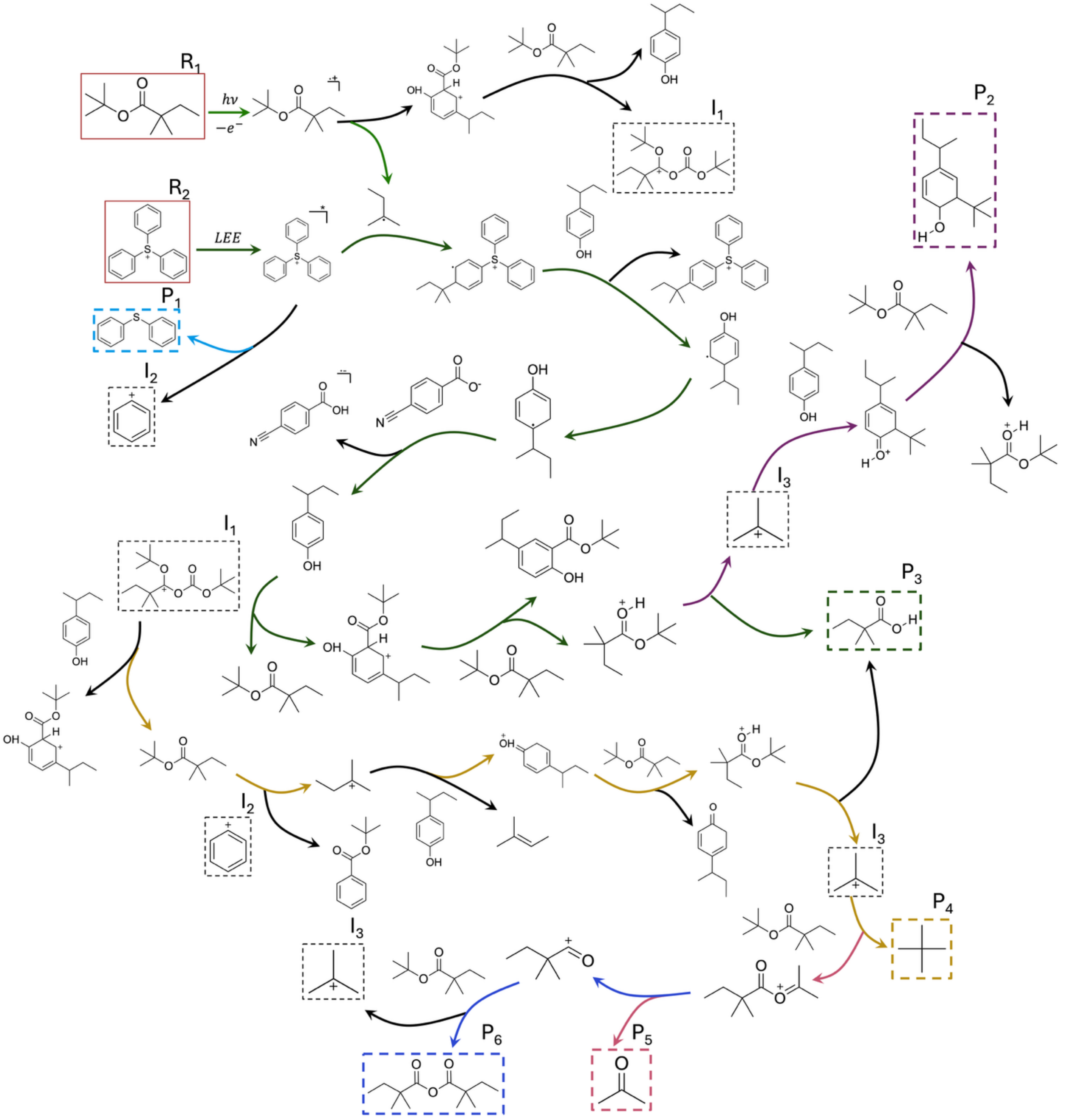

**Figure 7**. Kinetically significant reaction pathways following PtBMA photoexcitation, showing the sequential formation of key intermediates (dashed boxes) and products (solid boxes). The reaction sequences described above highlight the intermediates and elementary steps that connect the starting species to deprotection-related products and cross-linked molecular species. Arrows

are color coded by product-forming pathways, with each colored pathway leading to a distinct major product. Key reactants, intermediates and products are labelled as $R_n$, $I_n$, and $P_n$ respectively. Intermediates are shown in thin black dashed boxes, and products in colored dashed boxes.

The pathway analysis in Figure 7 shows how a broader set of products can be formed within the 5.15 $nm^3$ volume following photoexcitation of PtBMA. The deprotection reaction pathway is organized around cleavage of *tert*-butyl containing side chains to form carboxylic acid products, such as 2,2-dimethylbutanoic acid together with the corresponding alcohol and olefin fragments. Depending on whether PtBMA, PHS, Nf, TPS generate secondary electrons, or CNBZ is initially activated, the network reaches these intermediates through different combinations of proton transfer, electron transfer, hydrogen abstraction, and radical recombination steps. In contrast, the crosslinking pathway proceeds through coupling of long lived radical and cationic intermediates on the PHS and PtBMA backbones, generating new C–C and C–O connectivities that stitch neighboring chains and side groups together. These results show that EUV illumination creates a shared pool of radicals and ions that can produce either solubility-increasing deprotection/scission or solubility-reducing crosslinks, directly impacting image formation—i.e., competing outcomes from the same moiety, characterized as a result of methodological advancements introduced in this study.

Our simulations reproduce the established timescale hierarchy of EUV-driven radiolytic processes. Radical cation formation from photoionization is largely complete within the first ~100 fs. Over the following picoseconds, energetic and thermalized electrons drive both endergonic reactions (bond cleavage, molecular rearrangement) and exergonic reactions (electron trapping, bimolecular recombination) at and beyond the absorption site, with the full chemical evolution reaching completion on a timescale of tens of picoseconds to microseconds. This temporal ordering is consistent with prior experimental and computational studies of radiolysis in condensed organic media.[24, 40] The spatial distributions of products provide additional simulation validation: deprotection-related species are found within a ~10 nm domain surrounding the initial photoabsorption site, whereas crosslinked products are confined to within ~5 nm — spatial extents consistent with experimental and computational estimates of thermalized-electron travel distances and electron-driven product formation ranges in EUV resists.[22, 23, 40]

A central finding of this work is that the identity of the initially photoionized species profoundly shapes the downstream product distribution, through distinct sets of multi-step pathways rather than through variations in a single shared mechanism. When PtBMA or PHS absorbs the photon, ionic pathways dominate isobutylene formation via *tert*-butyl cation intermediates. When CNBZ absorbs, a largely radical-driven route produces isobutylene together with $CO_2$ without passing through an acid-mediated sequence. TPS, by contrast, does not drive local chemistry upon direct excitation but instead generates energetic electrons that activate reactions in the surrounding polymer matrix. These pathway differences have not been previously appreciated, and they challenge the conventional interpretation that isobutylene formation, the accepted experimental marker for deprotection, necessarily implies acid-catalyzed chemistry.[15, 48] Multiple mechanistic routes, both ionic and radical, can produce the same marker product, a conclusion accessible only through the kind of comprehensive, non-prescriptive mechanism discovery demonstrated here. EUV absorption cross sections are dictated by elemental composition, and absorbance scales with component abundance; since the polymer dominates the composition of photoresist films, polymer-centered absorption events can be expected to contribute substantially to the overall product distribution including extent of acid formation during exposure and its spatial characteristics, providing support for previous proposals in the literature.[26, 49]

Even within a ~5 $nm^3$ compartment volume, the simulations reveal a surprisingly broad chemical space: tens of distinct products generated through complex, kinetically favored, multi-step pathways that bear little resemblance to the one- or two-step mechanisms previously proposed for EUV resist chemistry. [22, 50-52] The competing formation of deprotection and crosslinking products from the same initial photoexcitation event, governed by branching among radical and ionic intermediates, provides a mechanistic basis for the experimentally observed dose-dependent transition from positive to negative tone behavior in similar resist systems.[46] Details such as the identity of intermediates, the branching ratios between competing pathways, and the spatial localization of each product class are inaccessible to approaches that prescribe reaction channels or lack spatiotemporal resolution. They emerge here because the workflow systematically constructs and filters the accessible reaction space rather than selecting pathways based on chemical intuition, and because the spatially resolved stochastic kinetic simulation tracks exactly where and when each product forms.

## Conclusions

Chemistry in concentrated condensed phases has not been examined at the level of mechanistic detail long established for gas-phase combustion and atmospheric systems, largely because the complexity of intermolecular interactions and the limitations of experimental detection at relevant length and time scales have made comprehensive mechanism characterization intractable. The integrated workflow described here, connecting high-throughput

DFT to automated reaction network construction, stochastic pathway sampling, and spatially resolved reaction–diffusion kinetics, demonstrates that detailed mechanistic information firmly rooted in physics and chemistry about how complex condensed-phase reactions evolve in space and time can be computationally accessed and validated against experiment. Applied to EUV-initiated radiolytic chemistry in a chemically amplified photoresist, the workflow reduced a space of millions of stoichiometrically possible reactions to approximately 8,300 kinetically relevant steps, produced spatiotemporal maps resolving product formation on femtosecond-to-nanosecond timescales across a 15.5-nm domain, provided detailed information on the nature of those products and predicted the formation of products that have been detected because of their volatility. We characterize that the identity of the initially absorbing resist component, and the distribution of secondary, low-energy, and thermalized electrons generated within the film, which determine the spatial and temporal distribution of the downstream chemistry after photoabsorption. The reactions forming these products are highly coupled and their kinetics evolve continuously with time, so that multiple key products (including deprotection and crosslinking) are generated concurrently from distinct precursors (i.e. initial absorbing components) and on different length scales from the initial absorbing site. We find that deprotection extends ~10 nm from the absorption site, while crosslinking remains confined within ~5 nm, so a nominally positive tone photoresist acquires negative tone character in the immediate vicinity of the initially deposited energy. Our results demonstrate that the efficiency and spatial fidelity of the latent image therefore emerge not from a single acid-generation step (as in DUV) but from coupled, time- and spatially-evolving reactions and products as a whole.

Several features of the workflow are essential to these results and distinguish it from prior computational approaches to condensed-phase mechanism discovery. The reaction enumeration and filtering proceed with fewer prescriptive assumptions than template-based methods, preserving a broader space of candidate pathways — including those that chemical intuition would not suggest. The coupling of filtering by chemical logic with stochastic pathway sampling identifies which reactions among millions of candidates actually participate in the chemistry. The spatially resolved kinetic simulation reveals not only what products form, but where they form relative to the energy deposition site and when they form relative to the initiating event. Such information is critical for connecting molecular-scale reactivity to the pattern-scale observables that determine photoresist performance and cannot be realized from any individual component of the workflow in isolation.

The current implementation has several limitations that define clear directions for improvement. The DFT calculations impose practical size limits on the molecular species that can be included; for example, the cation and anion components of TPS-Nf and TPS-CNBZ had to be treated separately rather than as closely associated ion pairs. DFT cost also precluded the inclusion of all possible fragments in recombination, such as those from C–C or C–F bond breakage in nonaflate. However, recent measurements on a related ESCAP-type photoresist observed release of HF and $CF_3$ as photoproducts, indicating that active pathways may have been excluded.[31] Condensed-phase species are modeled in an averaged solvation environment using an implicit solvent with a fixed dielectric constant, which may not capture the full complexity of local intermolecular interactions. Conformational diversity is not sampled during geometry optimization. These limitations can be substantially addressed by leveraging emerging machine learning interatomic potentials,[53-60] which would enable the inclusion of larger and more comprehensive molecular fragments and recombinants, explicit solvation environments, and conformational sampling at a fraction of the current DFT cost. Incorporation of computed reaction barriers, rather than the uniform rate coefficients used here, would further improve the physical accuracy of the kinetic simulations, particularly for reactions where barrier heights do meaningfully discriminate among competing pathways. The spatiotemporal outputs of the exposure-phase simulations described here can serve as inputs for subsequent reaction–diffusion modeling of the post-exposure bake process, enabling a complete computational connection from photoresist composition through latent image formation to developable latent image. Achieving this will require incorporating processes that are slow at ambient temperature but important during bake, such as diffusion of small neutral and protonated molecules through the polymer matrix and thermally activated processes. We note that careful management of the tradeoff between spatial resolution (smaller volume elements reduce kinetic bias from the well-mixed assumption) and computational cost will be required.

More broadly, the methodology established here provides a framework for quantifying complex reaction mechanisms with a level of mechanistic detail that has not previously been accessible for condensed-phase systems. The workflow is not specific to EUV lithography or photoresist chemistry. The same approach is applicable to any complex condensed-phase reactive system in which mechanisms are incompletely characterized, spatiotemporal resolution is needed to connect molecular-scale events to macroscopic observables, and experimental validation data are available. Examples include radiation-driven chemical transformations in soft matter, polymer degradation and curing processes, and interfacial reaction environments.

**ASSOCIATED CONTENT**

Data Availability Statement:

The data that support the findings of this study have been deposited in a GitHub repository at https://github.com/KMNitesh05/EUV_reactivity. The software used in this work is publicly available: HiPRGen and Reaction Network Monte Carlo (RNMC) are open-source packages, and Kinetiscope is freely available at http://www.hinsberg.net/kinetiscope.

**Supporting Information**. The Supporting Information is available free of charge at (DOI): network generation methodology, stochastic reaction-diffusion kinetic simulations, and spatiotemporal product distributions (PDF).

**AUTHOR INFORMATION**

Corresponding Authors

Samuel M. Blau – Energy Technologies Area**,** Lawrence Berkeley National Laboratory, Berkeley, CA 94720, United States *https://orcid.org/0000-0003-3132-3032*; Email: smblau@lbl.gov

Frances A. Houle − Chemical Sciences Division and Molecular Biophysics and Integrated Bioimaging Division, Lawrence Berkeley National Laboratory, Berkeley, California 94720, United States; *https://orcid.org/0000-0001-5571-2548*; Email: fahoule@lbl.gov

**Authors**

Nitesh Kumar – Materials Sciences Division, Lawrence Berkeley National Laboratory, Berkeley, CA 94720, United States; https://orcid.org/0000-0003-3322-8450

Jacob Milton – Materials Sciences Division, Lawrence Berkeley National Laboratory, Berkeley, CA 94720, United States

Eric Sivonxay – Energy Technologies Area, Lawrence Berkeley National Laboratory, Berkeley, CA 94720, United States

Brett A. Helms – The Molecular Foundry and the Materials Sciences Division, Lawrence Berkeley National Laboratory, Berkeley, CA 94720, United States

**Notes**

The Authors declare no competing financial interests.

**AI-Assisted Manuscript Preparation.** The authors used Claude (Anthropic) to assist with language editing. All AI-assisted edits were carefully reviewed, verified, and approved by the authors.

**ACKNOWLEDGMENTS**

This work was supported as part of the Center for High Precision Patterning Science (CHiPPS), an Energy Frontier Research Center funded by the U.S. Department of Energy, Office of Science, Basic Energy Sciences at Lawrence Berkeley National Laboratory under Contract No. DE-AC02-05CH11231. This work used computational resources provided by the National Energy Research Scientific Computing Center (NERSC), a U.S. Department of Energy Office of Science User Facility operated under Contract DE-AC02-05CH11231, and the Lawrencium computational cluster resource provided by the IT Division at the Lawrence Berkeley National Laboratory (Supported by the Director, Office of Science, Office of Basic Energy Sciences, of the U.S. Department of Energy under Contract No. DE-AC02-05CH11231). We thank Dr. William Hinsberg (Columbia Hill Technical Consulting) for helpful discussions.

TOC Figure

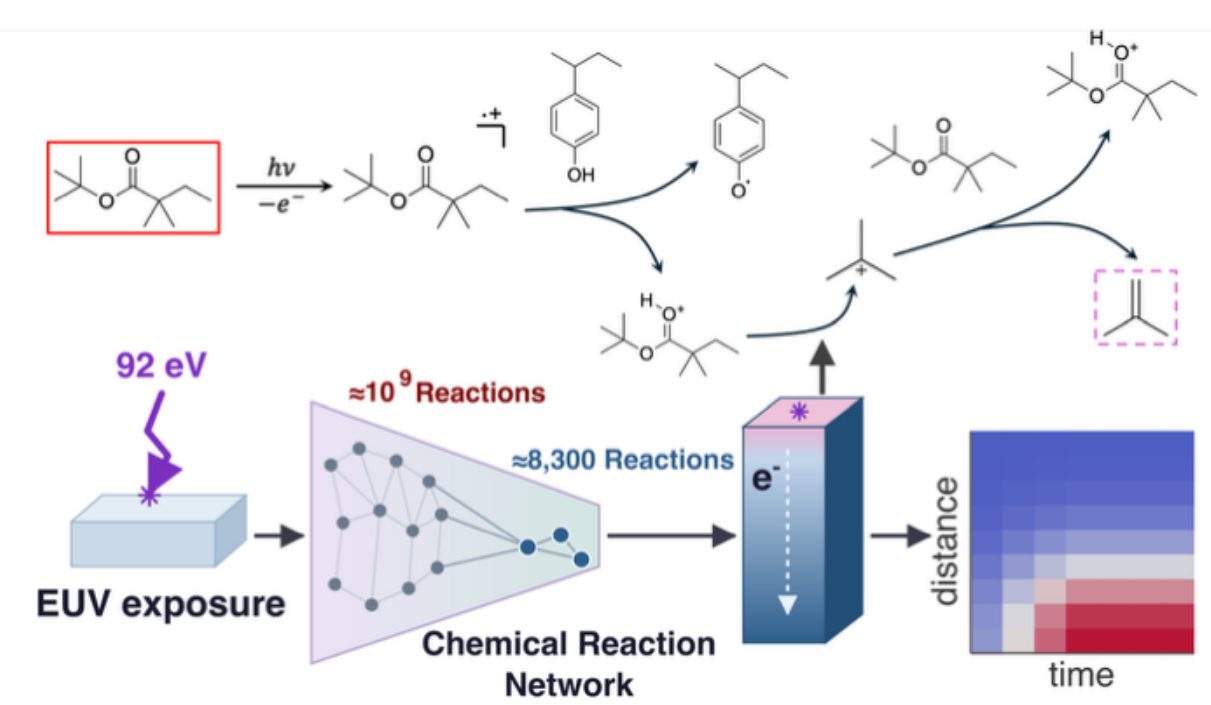

92 eV
EUV exposure
≈10⁹ Reactions
≈8,300 Reactions
Chemical Reaction Network
distance
time

# Supporting Information

# Discovering Kinetically Significant Reaction Mechanisms Beyond Chemical Intuition in Condensed-Phase Radiolysis

*Nitesh Kumar,[a,b] Jacob R. Milton,[a,b] Eric Sivonxay,[d] Brett A. Helms,[a,e] Frances A. Houle*[a,c,f] and Samuel M. Blau*[a,d]*

a. Center for High Precision Patterning Science, Lawrence Berkeley National Laboratory, Berkeley, California 94720, United States

b. Materials Sciences Division, Lawrence Berkeley National Laboratory, Berkeley, California 94720, United States

c. Chemical Sciences Division, Lawrence Berkeley National Laboratory, Berkeley, California 94720, United States

d. Energy Technologies Area, Lawrence Berkeley National Laboratory, Berkeley, California 94720, United States

e. The Molecular Foundry, Lawrence Berkeley National Laboratory, Berkeley, California 94720, United States

f. Molecular Biophysics and Integrated Bioimaging Division, Lawrence Berkeley National Laboratory, Berkeley, California 94720, United States

Email: smblau@lbl.gov, fahoule@lbl.gov

## Table of Contents

## Section S1. Network Generation Methodology

### Section S1a. Quantum Calculations for Structure Optimization

Quantum chemical density functional theory (DFT) calculations were carried out using version 6.1 of Q-Chem[1] with the SG3 integration grid[2] and our previously benchmarked high-throughput workflows.[3] Default settings were used except that the cutoff for neglecting two-electron integrals was tightened to $10^{-14}$ and molecular symmetry was disabled to avoid symmetry-breaking artifacts. These and related parameters were automatically adjusted when needed by the established error correction procedures.[3, 4] Geometry optimizations and vibrational frequency calculations employed the B97M-V functional[5] with the def2-SVPD basis set.[6, 7] Final single-point energies were refined using the ωB97M-V functional[5] with the def2-TZVPD basis.[6, 7] Solvent effects were modeled through the conductor-like polarizable continuum model (C-PCM)[8] using a dielectric constant of $\varepsilon = 3$, consistent with tabulated experimental values reported for poly(4-hydroxystyrene).[9] All optimized structures were confirmed to be true minima on the potential energy surface by ensuring the absence of imaginary frequencies, with the exception that a single very small imaginary mode (magnitude $\leq 15$ $cm^{-1}$) was accepted as numerical noise.

## Section S1b. HiPRGen Reaction Filtering Criteria

The reaction network for the ESCAP photoresist system is constructed in two stages that correspond to distinct physical regimes of EUV-initiated chemistry: radiolytic activation (Stage 1) and exergonic product-formation chemistry (Stage 2). The two-stage filter design reflects a physical separation: Stage 1 captures the endergonic, energy-driven chemistry of the radiolytic cascade (with the exception of exergonic electron attachment processes, which are physically inseparable from the ionization events that generate free electrons), while Stage 2 captures the exergonic product-formation chemistry that follows (with a +0.4 eV tolerance to account for systematic uncertainty in computed reaction energetics). Within each stage, filters are few in number, general, and physically motivated. They define what cannot reasonably occur — based on conservation laws, Coulombic constraints, spin selection rules, steric accessibility, and structural plausibility — rather than prescribing specific reaction types. This approach preserves a far broader reaction space than template-based enumeration would access.

Each stage applies a decision tree of sequential filters to every candidate reaction generated from the species set. A reaction enters the tree at the top and is evaluated against each filter in order. Some filters direct reactions into different branches of the decision tree, while others are terminal - i.e. if satisfied, cause the reaction to be kept or discarded. All reactions eventually encounter a terminal filter. This structure ensures that every candidate reaction follows a single, deterministic path through the filter logic.

Both stages share a common structural viability test, referred to here as fragment matching. For each candidate reaction, the algorithm attempts to decompose the reactant and product molecules into fragments and verify that the fragments on both sides can be related by breaking at most one bond in the reactants and breaking at most one bond in the products (corresponding to a bond formation as part of the reaction). This enforces that all surviving reactions correspond to elementary or near-elementary steps — concerted processes involving the transfer, loss, or gain of a single molecular fragment — rather than multi-step transformations compressed into a single reaction. Reactions for which no valid fragment decomposition exists are discarded.

Both stages also employ a local environment comparison (star count difference, defined below) as a coarse filter for the extent of bonding rearrangement. For each atom in a molecule, a graph hash of its immediate bonding neighborhood (the atom and all atoms covalently bonded to it) is computed. The star count difference between reactants and products is the total number of local environments that change across the reaction. Higher values indicate more extensive bonding rearrangement. Thresholds of 4 and 6 are used in different contexts: a threshold of 4 permits reactions in which at most one bond is broken (appropriate for unimolecular fragmentations), while a threshold of 6 permits reactions in which one bond is broken and one bond is formed (appropriate for bimolecular fragment-transfer reactions). This step precedes fragment matching because it is substantially less computationally expensive, dramatically reducing the number of reactions subjected to the more expensive fragment matching procedure.

**Stage 1: Radiolytic Activation**

Stage 1 captures the energy-driven processes initiated by the ~92 eV deposited by an EUV photon: photoionization, fragmentation of radical cations and electronically excited species, and endergonic transformations accessible within the deposited energy budget. Electron attachment is also captured in this stage. The filters are organized into three branches based on reaction type.

**Redox reactions** (change in total charge between reactants and products)

Redox reactions — those in which the total formal charge changes between reactants and products — are processed first. To be retained, a redox reaction must satisfy all of the following:

- It involves exactly one reactant and one product (unimolecular).

- The change in formal charge is exactly ±1 (single electron transfer).

- The reactant and product have identical molecular graphs (covalent connectivity is preserved; only the charge state changes).

Reactions satisfying these criteria are further divided into electron attachment (charge becomes more negative) and ionization (charge becomes more positive). For electron attachment reactions, two additional filters apply. Attachment to species that are already negatively charged is discarded, reflecting the Coulombic barrier that makes such processes far slower than attachment to neutral or positively charged species. Additionally, attachment reactions are required to have an adiabatic free energy change ($\Delta G$) more negative than −0.44 eV. This threshold was determined by correlating computed adiabatic $\Delta G$ values with vertical electron affinities across the species set; −0.44 eV is the adiabatic $\Delta G$ at which the vertical electron affinity crosses zero. Species with less favorable adiabatic $\Delta G$ have negative vertical electron affinities, meaning the electron would not be captured on the timescale of the attachment event — the geometric relaxation that makes the adiabatic process exergonic would not have time to occur. Ionization reactions (electron loss) are kept with no thermodynamic threshold, consistent with the energy-driven nature of the radiolytic activation process. Note that exergonic electron attachment reactions are the only exergonic reactions contained in Stage 1.

**Bimolecular non-redox reactions (two reactants)**

Non-redox reactions with two reactants are heavily constrained in Stage 1. Candidate reactions are discarded if they meet any of the following criteria:

- Only one product is formed (association reactions).

- Both reactants carry the same sign of charge (Coulombic repulsion prevents approach).

- Both reactants are closed-shell singlets and both products are open-shell (radicals would immediately recombine in the solid polymer matrix).

- The reaction separates charge (products carry charges of larger magnitude than reactants, or two neutral reactants produce charged products).

- A spectator species is present: one molecular graph appears on both sides of the reaction with the same charge state, meaning the reaction is effectively unimolecular.

- The reaction involves simultaneous electron transfer and fragment transfer (coupled electron–fragment transfer), unless fragment matching validates the structural rearrangement.

- The star count difference exceeds 6 (more than one bond broken and one formed).

- The compositions of reactants and products are inconsistent with single hydrogen transfer (the only bimolecular fragment transfer permitted in Stage 1).

Bimolecular reactions that survive these filters must additionally pass fragment matching and must be endergonic ($\Delta G > 0$), consistent with the requirement that Stage 1 reactions consume deposited energy. Two further filters refine the type of hydrogen transfer permitted. Neutral H-atom (H•) abstraction from a closed-shell reactant — that is, abstraction by a radical, R• + R′–H → R–H + R′• (abstraction by another closed-shell species is already removed by the two-radical-product filter above) — is discarded from Stage 1. Such radical-mediated abstractions are thermally driven propagation steps rather than energy-consuming activation events: their thermodynamically favorable direction is captured as fragment transfer in Stage 2, whereas the only versions that could satisfy the Stage 1 $\Delta G > 0$ criterion would convert a more stable radical into a less stable one and are not expected to compete with the reverse process. Hydride ($H^-$) abstraction is also discarded from Stage 1. Hydride transfer between two neutral species separates charge and is already removed by the charge-separation filter; hydride abstraction by a cation ($R–H + C^+ \rightarrow R^+ + C–H$) is feasible when the newly formed cation is more stable than the abstracting one, but in that case the transfer is exergonic and thermally driven, and so falls outside the endergonic, energy-consuming reaction set that Stage 1 is constructed to capture. The surviving bimolecular reactions are therefore H-atom transfers from open-shell (radical) species and proton ($H^+$) transfers from any species.

**Unimolecular non-redox reactions (single reactant)**

Non-redox reactions with a single reactant must be endergonic ($\Delta G > 0$; exergonic reactions are explicitly discarded). Single-reactant, single-product isomerizations are discarded given that they are highly unlikely to occur in a rigid polymer matrix at ambient temperature. The star count difference threshold is set to 4, permitting only single-bond cleavage events. Reactions that produce two radical products from a closed-shell reactant (homolytic radical separation) and reactions that separate charge are both discarded. Surviving reactions must pass fragment matching, and ring-closing reactions (where one of two products results from intramolecular bond formation in the reactant) are discarded.

**Stage 2: Exergonic Chemistry**

Stage 2 captures the exergonic processes through which the reactive intermediates produced during radiolytic activation evolve toward stable products: fragment transfer, radical recombination, charge transfer, and molecular rearrangement. The filters are substantially simpler than in Stage 1.

All redox reactions are discarded — Stage 2 contains no ionization or electron attachment processes. Reactions are filtered by a thermodynamic criterion that retains exergonic reactions and those that are endergonic by no more than +0.4 eV ($\Delta G \leq +0.4$ eV). This tolerance accounts for the combined systematic uncertainty in computed reaction free energies arising from the density functional approximation, implicit solvation (which does not capture the full dielectric and hydrogen-bonding environment of the dense polymer matrix), the use of truncated molecular fragments rather than full polymer chains to represent the backbone, and the neglect of conformational sampling. Each of these approximations introduces an error of modest magnitude individually, but their combined effect can shift computed reaction energies by several tenths of an eV in a consistent direction, such that reactions computed to be mildly endergonic may in fact be thermoneutral or exergonic in the real system. Reactions between two species of the same charge sign are discarded.

Charge transfer reactions — in which the molecular graphs of reactants and products are identical but the charge states differ (e.g., $A^+ + B \rightarrow A + B^+$)—are explicitly retained. Reactions containing a spectator species (same molecular graph and charge on both sides) and coupled electron–fragment transfer reactions are discarded. The star count difference threshold is 6.

All remaining reactions must pass fragment matching. Within the fragment-matched set, three additional filters apply:

- Single-reactant, single-product reactions that are not intramolecular hydrogen or fluorine atom transfers are discarded (blocking non-atom-transfer isomerizations).

- Ring-closing reactions are discarded.

- Sterically hindered reactions are discarded. A reaction is classified as hindered if the atoms at both the bond-breaking and bond-forming sites are sp3-hybridized carbons with three or more sp3 carbon neighbors (tertiary or quaternary centers), unless at least two of the reactive atoms are methyl carbons. This filter reflects the physical constraint that bulky substituents at the reactive centers prevent the close approach required for bond formation in a dense polymer matrix.

For reactions that pass fragment matching but involve a spectator molecular graph (same covalent connectivity appearing on both sides, potentially with different charge states), an additional filter retains only those whose fragment decomposition has the pattern A + A + B (two identical fragments plus one distinct fragment). This permits bimolecular reactions such as radical disproportionation where one equivalent of a species is consumed and one is regenerated, while discarding reactions where the spectator is truly uninvolved.

## Section S1c. Species and Reaction Naming Convention for Kinetiscope Input

The chemical reaction network (CRN) generated by HiPRGen[10] identifies each molecular species by a unique string composed of an MD5 hash, the molecular formula, the formal charge, and the spin multiplicity (e.g., 0086d2a354484d6f2ef1cba0fd711790-C16H19S1-0-2). While these identifiers are unique and suitable for internal database operations, they are neither human-readable nor compatible with the naming requirements of the Kinetiscope kinetic simulation package.[11] We therefore implemented an automated renaming pipeline to convert all species identifiers and reaction strings into a compact, self-documenting format that encodes the essential electronic-structure metadata required for kinetic modeling.

### Renaming Procedure

The renaming pipeline operates in three stages. First, the HiPRGen JSON output containing HiPRGen_Reaction objects is parsed recursively to extract all unique species appearing as reactants or products across the full reaction network. Second, each species identifier is parsed via regular expression matching to extract four fields: (i) the MD5 hash (discarded), (ii) the molecular formula, (iii) the formal charge $q$, and (iv) the spin multiplicity $2S + 1$. The total electron count $N_e$ is computed from the molecular formula and formal charge using tabulated atomic numbers (H = 1, C = 6, N = 7, O = 8, F = 9, S = 16):

$$N_e = \Sigma_i Z_i n_i - q$$

where $Z_i$ is the atomic number and $n_i$ is the number of atoms of element $i$. A radical/charge descriptor tag is assigned according to the electronic state: r for neutral radicals (spin multiplicity

> 1, $q = 0$), rc for radical cations (spin multiplicity > 1, $q = +1$), ra for radical anions (spin multiplicity > 1, $q = -1$), and no tag for closed-shell species. Each species is further assigned a unique integer index $M_n$ ($n = 1, 2, \ldots$) based on the lexicographic order of the original identifiers, ensuring one-to-one correspondence between the HiPRGen database entries and the Kinetiscope species names.

The resulting species name follows the general format:

{formula}_{charge}__{Ne}_{tag}_M{index}

Third, each reaction in the HiPRGen network is reformatted by replacing all reactant and product identifiers with their corresponding Kinetiscope names. The reaction string is constructed according to Kinetiscope format, where reactants and products have appropriate stoichiometries and are connected with species joined by " + " on each side of an " => " delimiter, where the delimiter signifies a non-reversible arrow connecting reactants to products. The full list of reactions, which constitute the mechanism to be simulated, is exported alongside reaction metadata as a comma-separated file for direct import into Kinetiscope.

**Worked Example**

To illustrate the naming convention, consider the species with HiPRGen identifier 0086d2a354484d6f2ef1cba0fd711790-C16H19S1-0-2. Parsing this string yields the molecular formula $C_{16}H_{19}S_1$, formal charge $q = 0$, and spin multiplicity 2 (doublet). The total electron count is:

$$N_e = (16 \times 6) + (19 \times 1) + (1 \times 16) - 0 = 131$$

Since the spin multiplicity exceeds unity and the formal charge is zero, the species is classified as a neutral radical and assigned the tag r. As the first entry in the sorted species list, it receives the index M1. The complete name for the kinetics scheme is therefore:

C16H19S1_0_2_131_r_M1

Table S1 summarizes the field definitions encoded in this name.

**Table S1.** Decomposition of the Kinetiscope species name C16H19S1_0_2_131_r_M1 into its constituent fields.

| Field | Value | Description |
|---|---|---|
| Formula | C16H19S1 | Molecular formula (element–count pairs) |
| Charge | 0 | Formal charge (0 = neutral; m prefix encodes negative, e.g., m1 = −1) |
| Spin multiplicity | 2 | 2S + 1 (1 = singlet, 2 = doublet, 3 = triplet) |
| Electron count | 131 | Total electron count $N_e = \Sigma Z_i n_i - q$ |
| Tag | r | Electronic state descriptor: r = neutral radical, rc = radical cation, ra = radical anion, none = closed-shell |
| Index | M1 | Unique molecule index (lexicographic order of original HiPRGen identifiers) |

This naming convention ensures that each species in the Kinetiscope simulation retains its essential electronic-structure metadata—formula, charge state, spin multiplicity, electron count, and radical character—in a format that is both compact and immediately interpretable during kinetic analysis. A representative cation fragmentation reaction in the reformatted notation reads:

C18H15S1_1_1_138_M1162 => C12H10S1_0_1_98_M94 + C6H5_1_1_40_M870

where the reactant is a closed-shell cation (charge +1, singlet, no radical tag) fragmenting into a neutral closed-shell product and a cation fragment, with each species fully characterized by its name alone.

**Initial Photoexcitation and Radiation Physics**

The reaction scheme used to describe the initial ionization, excitation and electron trapping events has been described in detail elsewhere[12] and briefly summarized here. The steps involved for all 5 resist components X are of the following form:

Photoabsorption:

X + EUV photon => X+ + 80 eV electron + LEE

Ionization:

X + 80 eV electron => X+ + 55 eV electron + LEE

X + 55 eV electron => X+ + 30 eV electron + LEE

X + 30 eV electron => X+ + 2 LEE

Electronic excitation:

X + LEE => X* + TE

Electron trapping:

TE + X+ => X

TE + X => X-

Electronic deexcitation:

X* => X

where LEE are low energy electrons (~ 5 eV), TE are thermal electrons (0.1 eV), and electron trapping to form an anion only occurs when the species has a finite electron affinity. Rate coefficients for each step were calculated from absorption cross sections, times for electrons of a given energy to move the scattering distance of 0.79 nm,[13] or the Smoluchowski or reduced Debye equations for diffusion controlled electron trapping. Electronic deexcitation was assumed to be from the triplet manifold because radiolysis is known to form abundant triplet populations, with a lifetime of 0.5-1 ms. Electron diffusion rate coefficients were obtained by calculating the time an electron of a specific energy takes to travel 1 nm. Any of the products, X+, X* or X-, can undergo chemical reactions competitive with electron-molecule processes as defined by the reaction network. TPS is the only species in the scheme that begins as a positive ion and therefore forms a doubly charged cation during the ionization steps. Accordingly, it has an extra TE trapping step TE + X2+ => X+.

## Section S2. Stochastic Reaction Diffusion Kinetic Simulations

The reaction scheme resulting from the above creation process is directly imported into Kinetiscope, a general purpose, open access stochastic kinetics simulation package.[11] The core algorithm treats the chemical reaction as a Markov process, and generates an exact solution to the master equation for the system at all points in time. This method has the advantage of avoiding stiffness common to ordinary differential equation solvers, which often can only be managed via changes to the mechanism. Use of value for k ($k_bT/h = 6.2 \times 10^{12}$ ) for both first and second order reaction steps in the mechanism overestimates the second order reaction rate coefficients by about a factor of 10, however the error introduced by this simplification for a relatively small number of steps is negligible in a mechanism of this size. The simulation engine uses the scheme, the initial concentrations and the rate coefficients for each step to calculate instantaneous rates for each step in the scheme. In the present study, movement of electrons between compartments in the 1-D array occurs via transfer paths that have associated diffusion rate coefficients. In the simulation loop, instantaneous values for the reaction and transfer rates are normalized and equated to probabilities, and a random number is used to select among the probability-weighted reaction steps. The sum of the rates is used to calculate the magnitude of the time step associated with the occurrence of the selected step. Following an update to concentrations of reactants and products according to the step stoichiometry or the transfer direction, rates (and therefore probabilities) are recalculated, a new random number is obtained, and a new selection cycle begins. The simulation ends at either a pre-specified time point, or at the time when all reaction rates are zero. The latter criterion was used in the present work. The output of the simulation is a complete history of all species in the system in space and time that can be analyzed to provide a deeper understanding of the mechanism itself.

## Section S3. Construction of Averaged Spatiotemporal Distributions from Kinetiscope Output

Kinetiscope[11] performed stochastic kinetic simulations on a one-dimensional spatial grid, producing species amounts as a function of both time and distance from the compartment where photoabsorption takes place. For each species of interest, the simulation output is exported as an

eXtensible Model Format (XMF) file—an XML-based scientific data format—that encodes the spatially resolved amount (in moles) at each recorded time step. Because Kinetiscope employs a stochastic algorithm launched by a random number seed,[14, 15] individual simulations exhibit run-to-run statistical fluctuations. To obtain spatiotemporal profiles with less random fluctuation, we performed $N_{rep} = 5$ independent replicate simulations for each system using different random number seeds and averaged the results on a common temporal grid, as described below.

**XMF File Structure and Parsing**

Each Kinetiscope XMF export file is structured as an XML document containing a list of the amounts of each species in each compartment at successive time points. In the present simulations, the one-dimensional spatial domain spans 15.5 nm with the photoabsorption compartment being defined to be the first one, and is discretized into $n_c = 31$ uniform compartments of width $\Delta x = 0.5$ nm each. Parsing a single XMF file therefore yields a time vector $\mathbf{t}$ of length $n_t$ and a two-dimensional data array $\mathbf{A}$ of shape ($n_c \times n_t$), where $A_{jk}$ is the amount (mole) of the species in compartment $j$ at time $t_k$.

**Temporal Interpolation and Run Averaging**

Because the stochastic algorithm in Kinetiscope records snapshots at internally generated time points,[14, 15] the time grids from different replicate runs will differ slightly in the number or spacing of recorded points. To place all replicates on a common footing, we adopted the time grid of the first replicate, $\mathbf{t}^{ref}$, as the reference. For each subsequent replicate $i$ ($i = 2, \ldots, N_{rep}$), the data array $\mathbf{A}^{(i)}$ was interpolated onto $\mathbf{t}^{ref}$ by one-dimensional linear interpolation applied independently to each compartment $j$:

$$\tilde{A}_{jk}^{(i)} = \mathrm{interp}(t_k^{ref}; \mathbf{t}^{(i)}, \mathbf{A}_j^{(i)})$$

where the time points of replicate $i$ were first sorted in ascending order to ensure monotonicity. The run-averaged spatiotemporal distribution was then computed as the arithmetic mean over all replicates:

$$\langle A_{jk} \rangle = (1/N_{rep}) \Sigma_i \tilde{A}_{jk}^{(i)}$$

The standard deviation associated with the arithmetic mean so obtained is $\pm \sigma_{jk} / \sqrt{N_{rep}}$, where $\sigma_{jk}$ is the sample standard deviation of the interpolated replicate values $\tilde{A}_{jk}^{(i)}$ at each compartment–time point (j, k).

### Coordinate Construction and Visualization

Compartment edges were constructed as $n_c + 1$ uniformly spaced values from 0 to $n_c \times \Delta x$ (= 15.5 nm) as defined in the simulation framework. Temporal bin edges were placed at the midpoints between consecutive recorded times, with half-interval extensions at the first and last time points. Time was converted from seconds to femtoseconds (1 fs = $10^{-15}$ s) for visualization.

The averaged spatiotemporal distributions $\langle A_{jk} \rangle$ were visualized as two-dimensional heatmaps with time on the horizontal axis and distance from the exposure site on the vertical axis, with the color scale representing the species amount in moles. In selected cases, the raw amount was converted to a number density (mole nm$^{-3}$) by dividing by the compartment volume $V_{bin}$ = 5.15205 nm$^3$. Additionally, one-dimensional time traces of the amount in the photoabsorption compartment ($j$ = 1, corresponding to 0–0.5 nm from the outer edge) were extracted to visualize the temporal evolution of species accumulation there.

### Extraction of Time-Resolved Reaction Pathways from Kinetiscope Time-Sequence Data

Kinetiscope uses a particle-based algorithm, where each particle represents a defined amount of matter. In the present work, each molecular species is represented by one particle, an amount of $1.66 \times 10^{-24}$ moles. Because of the photoresist composition and compartment volume, each compartment contains 25 particles that are well mixed at all times. Kinetiscope records the full state of the system after every selection, generating the complete chronological sequence of selected reactions. This step-by-step record provides the raw data from which time-resolved reaction pathways, the ordered sequences of elementary reactions connecting initial species to any target product, can be reconstructed. The pathway extraction procedure is described below and illustrated with a worked example.

### Construction of the Time-Resolved Step Selection Log

In addition to the time histories of all chemical species, Kinetiscope generates a matrix in which each row corresponds to a recorded time point and each column corresponds to a reaction step in the scheme. The matrix entries are cumulative counts of how many times each step has been selected up to that time, enabling the provenance of various products to be traced unambiguously. This table is first converted to a per-interval increment representation by computing the forward difference along the time axis, yielding the number of times each step was selected within each time interval.

The incremented counts are then collapsed into a compact event log: for each time point, the set of step numbers with nonzero increments is collected.

**Time-Resolved Pathway Tracing to a Target Product**

Given a target product species, identified by its unique molecule index (e.g., M383 for 2,2-dimethylbutanoic acid, $C_6H_{12}O_2$), the time-resolved pathway is traced by a forward-sweep algorithm that links each consumed reactant to its most recent producer in the reaction step selection log. The algorithm operates as follows:

(i) The full reaction scheme is loaded as a lookup table mapping each step number to its reactant and product species lists, parsed from the reaction equation strings. Markers (e.g., excite, LEE, ionization, eV_80, used to track totals of types of events)[12] are excluded from the chemical species sets.

(ii) A production event index is built by scanning all selected reaction steps in the log chronologically: for each selected step, every product species is recorded under both its full token name and its M-tag (the unique molecule index extracted by the pattern M{integer}), creating a mapping from each species to the sorted list of (time, step) pairs at which it was produced.

(iii) Starting from the earliest event in the event selection log, the algorithm sweeps forward in time. At each event, every reactant is checked against the production event index to find whether it was produced by an earlier event in the pathway being constructed. If a reactant has no prior producer within the pathway but does have one in the global production index, that producer event is inserted into the pathway (augmentation). This ensures that the chain captures all intermediate transformations, even if they were not immediately adjacent in time.

(iv) The sweep terminates when a reaction is encountered whose products include the target species (matched by M-tag). The resulting pathway is a time-ordered list of (time, step number, reaction equation) tuples representing the minimal time-resolved chain from initial excitation or seed species to the target product.

**Provenance Table and Dependency Graph**

For each step in the extracted pathway, a provenance table is constructed that records, for every reactant consumed, the pathway index, step number, and time at which that reactant was produced by an earlier step. Reactants with no identified producer (seed species) are marked with null entries. This table enables direct inspection of how intermediate species are passed from one reaction to the next, and which species are recycled through loops within the pathway.

The provenance relationships are also represented as a directed acyclic graph in which nodes correspond to selected reaction events and edges encode two types of relationships: (a) chronological chain edges connecting temporally adjacent events, and (b) dependency edges linking a producer event to a consumer event via the shared intermediate species (labeled by its M-tag). The graph is rendered with Graphviz and, optionally, as an interactive PyVis visualization. In the species-centric view, a bipartite graph is constructed in which reaction nodes (rectangles) alternate with species nodes (ovals, optionally annotated with molecular structure images), making the flow of chemical matter through the pathway visually explicit.

**Worked Example: Pathway to 2,2-Dimethylbutanoic Acid (M383)**

The time-resolved pathway to M383 ($C_6H_{12}O_2$, 2,2-dimethylbutanoic acid) was extracted from a Kinetiscope simulation of the reaction sequence initiated by EUV photoexcitation of poly(*tert*-butyl methacrylate) (PtBMA). The algorithm identified an 11-step time-resolved chain traversing 323 time steps (~26.7 ps). Table S2 lists the pathway in chronological order, and the provenance of each reactant is annotated.

**Table S2.** Time-resolved reaction pathways to the formation of 2,2-dimethylbutanoic acid (M383) extracted from the sequential selection of reactions during stochastic reaction diffusion simulation. The “Bridge species” column indicates which intermediate species links each reaction to a prior step in the chain; “seed” denotes species present at the start of the simulation.

| # | Time Step Number | Reaction | Bridge | Type |
|---|---|---|---|---|
| 1 | 1 | M1334 + excite → M1340 + eV_80 | seed | Ionization |
| 2 | 6 | M840 + M1340 → M1922 + M2411 | M1340 ← step 1 | Radical add. |
| 3 | 16 | M2148 + LEE → M2148* | seed | Excitation |
| 4 | 78 | M2148 + M2411 → M2969 | M2411 ← step 2 | Radical comb. |
| 5 | 83 | M840 + M2969 → M1408 + M1850 | M2969 ← step 4 | H-transfer |
| 6 | 202 | M1334 + M1922 → M840 + M1870 | M1922 ← step 2 | Proton transfer |
| 7 | 235 | M1408 → M1407 | M1408 ← step 5 | Isomerization |
| 8 | 282 | M1407 + M36 → M840 + M1223 | M1407 ← step 7 | Electron transfer |
| 9 | 299 | M840 + M1870 → M1334 + M1922 | M840 ← step 8 | Proton transfer |
| 10 | 300 | M1334 + M1922 → M775 + M166 | M1334, M1922 ← step 9 | Charge transfer |
| 11 | 323 | M775 → M2162 + M383 | M775 ← step 10 | Fragmentation |

The pathway begins with the ionization of a PtBMA repeat unit (M1334) at time step number $t$ = 1, producing a radical cation (M1340). This radical cation undergoes a bimolecular addition with a neutral ester fragment (M840) at $t$ = 6 (step 2), generating two key intermediates: M1922 and a pentyl radical M2411. In parallel, the photoacid generator (M2148, triphenylsulfonium) is electronically excited at $t = 16$ (step 3). The pentyl radical M2411 produced in step 2 then combines with the excited PAG at $t = 78$ (step 4), forming a large radical cation adduct M2969.

Steps 5–9 describe a cascade of hydrogen-atom transfers, isomerizations, and electron-transfer reactions that regenerate M840 and M1334 through loops—a notable feature revealed by the provenance analysis. Specifically, M840 is consumed in step 5 and regenerated in step 8, while M1334 is consumed in step 6 and regenerated in step 9, indicating that these species act as transient carriers rather than being permanently consumed. Step 10 converts M1334 and M1922 into a protonated ester cation M775, which finally undergoes unimolecular fragmentation at $t = 323$ (step 11) to yield the target product M383 (2,2-dimethylbutanoic acid) along with an isobutyl cation M2162.

This example demonstrates how the step-by-step recording of Kinetiscope event selection steps, combined with the time-resolved tracing algorithm, reveals not only the sequence of elementary reactions leading to a specific product starting with a specific random number seed, but also the catalytic recycling of intermediate species and the coupling between independent initiation events (photopolymer ionization and PAG excitation) that would be invisible in conventional concentration-vs-time profiles.

## Section S4. Spatiotemporal distributions of other key experimentally observed products

The spatiotemporal distributions of additional experimentally detected small molecule products are presented in Figure S2. These species — including water (a), acetone (b), benzene (c), $CO_2$ (d), sec-butylbenzene (e), tert-butylbenzene (f), n-butane (g), and isobutane (h) — exhibit distinctly different spatial signatures from the deprotection and crosslinking products in Figure 4 in the main paper. Notably, aromatic molecules like benzene and tert-butylbenzene appear only at > 2 nm from the absorption site, indicating that these species are formed exclusively through secondary electron-driven reactions in adjacent compartments rather than by direct fragmentation at the point of photon absorption. Water is the most spatially confined, concentrated primarily within 3 nm of the absorption site and persistent throughout the simulation timescale. In contrast, $CO_2$ and acetone appear more sporadically and predominantly at later times (>3000 fs), consistent with their formation through multi-step fragmentation pathways that require the prior generation of intermediate radical or ionic species. The aromatic fragments — benzene, sec-butylbenzene, and tert-butylbenzene — display banded spatial distributions extending out to 10–15 nm, reflecting the range of thermalized and low-energy electrons that initiate bond scission in the polymer matrix. These results confirm that several products previously detected by mass spectrometry during EUV exposure originate from chemistry occurring well outside the immediate photoabsorption volume, underscoring the role of electron transport in determining the spatial extent and compositional heterogeneity of the latent image.

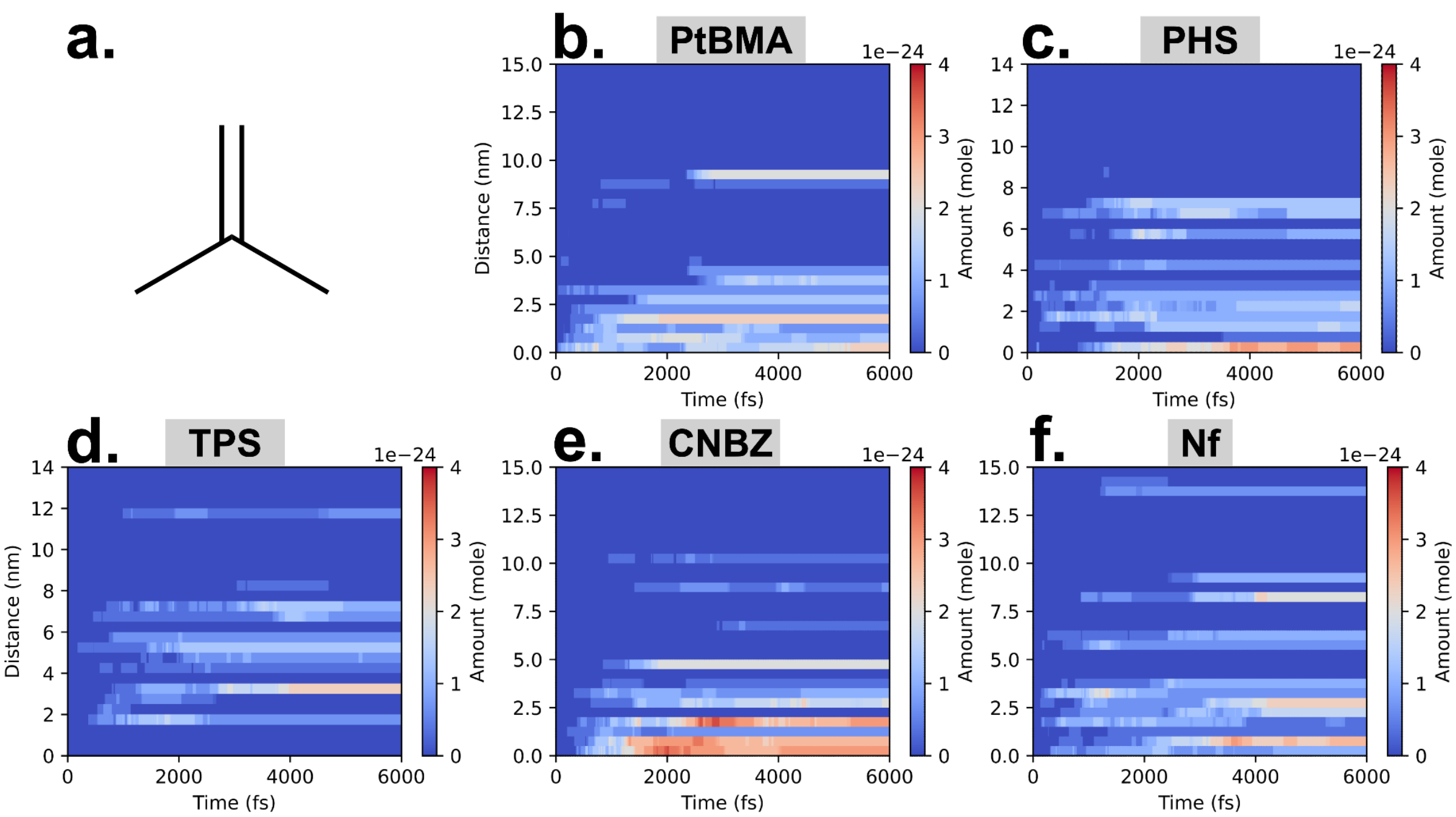

**Figure S1.** Spatiotemporal distributions of (a.) isobutylene for each initially excited component ((b.) PtBMA, (c.) PHS, (d.) TPS, (e.) CNBZ, (f.) Nf respectively) of the photoresist, plotted as amount (moles, where $1.66 \times 10^{-24}$ moles is equal to one molecule) versus time (up to 6000 fs) and distance (nm). Amounts are reported as averages over 25 independent simulations.

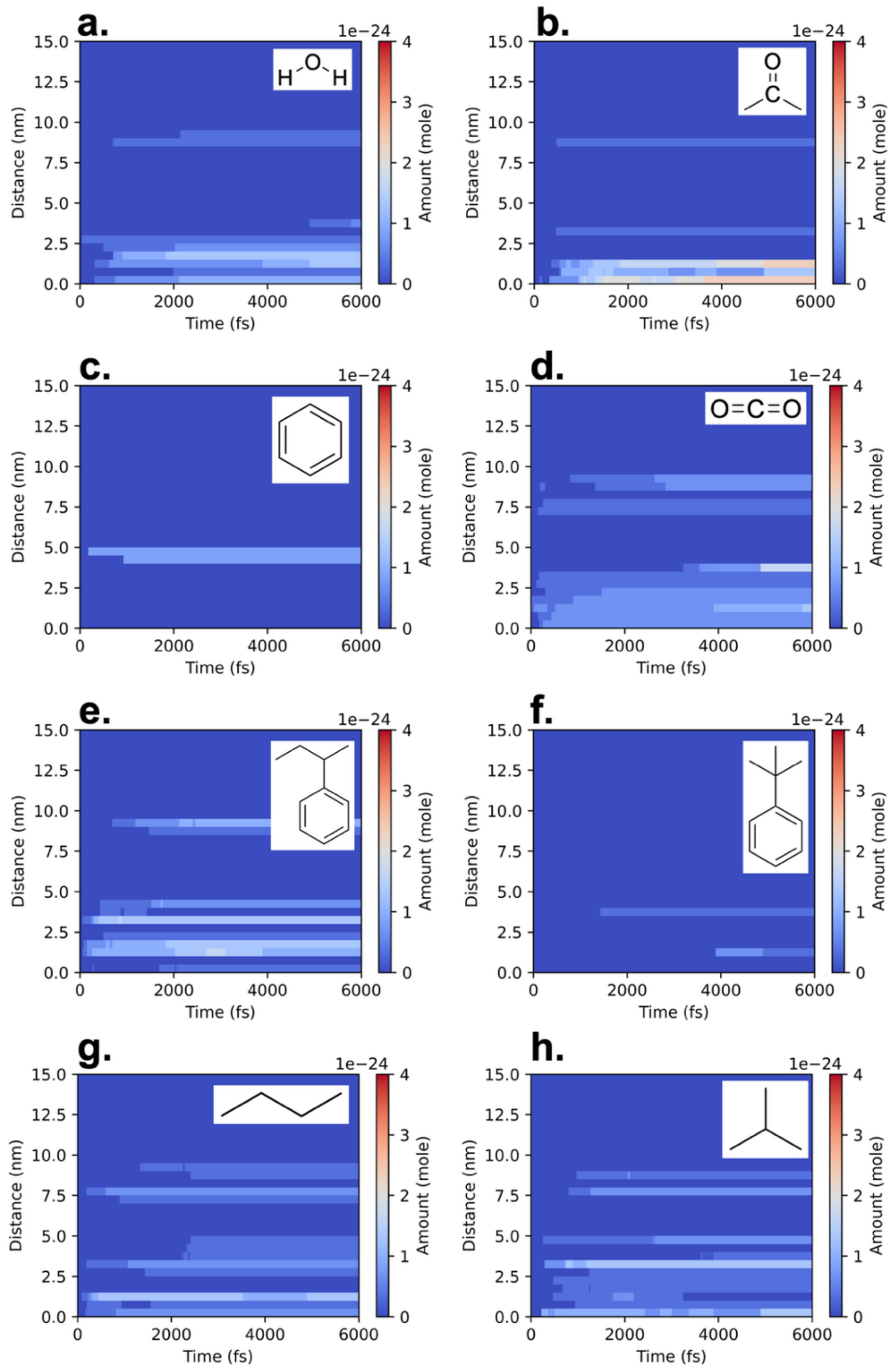

**Figure S2.** Spatiotemporal distributions of experimentally detected small molecule products: **(a)** water, **(b)** acetone, **(c)** benzene, **(d)** carbon dioxide, **(e)** sec-butylbenzene, **(f)** tert-butylbenzene, **(g)** n-butane, and **(h)** isobutane, plotted as amount (moles, where $1.66 \times 10^{-24}$ moles is equal to one

molecule) versus time (up to 6000 fs) and distance (nm) from the initial EUV absorption site (compartment spanning 0–0.5 nm). Amounts are reported as averages over 25 independent simulations. Insets show the molecular structure of each species.